\documentclass[sigconf,balance=false]{acmart}
    
\usepackage{tabularx}
\usepackage{booktabs}

{
  \color{black}
  \begin{itshape}%
  \doublespacing
}%
{\end{itshape}}

\copyrightyear{2023}
\acmYear{2023}
\setcopyright{rightsretained}
\acmConference[CHI '23]{Proceedings of the 2023 CHI Conference on Human Factors in Computing Systems}{April 23--28, 2023}{Hamburg, Germany}
\acmBooktitle{Proceedings of the 2023 CHI Conference on Human Factors in Computing Systems (CHI '23), April 23--28, 2023, Hamburg, Germany}\acmDOI{10.1145/3544548.3580833}
\acmISBN{978-1-4503-9421-5/23/04}

\newcommand{\revise}[1]{{\textcolor{black}{{#1}}}}

\newenvironment{revision}
  {\color{black}}
{}

\newcommand{\revis}[1]{{\textcolor{black}{{#1}}}}

\newenvironment{revisio}
  {\color{black}}
{}

\begin{document}

\title[Supporting Piggybacked Co-Located Leisure via AR]{Supporting Piggybacked Co-Located Leisure Activities via Augmented Reality}


\author{Samantha Reig$^1$ \quad Erica Principe Cruz$^1$ \quad Melissa M. Powers$^2$ \quad Jennifer He$^3$ \\ Timothy Chong$^4$ \quad Yu Jiang Tham$^5$ \quad Sven Kratz$^6$ \quad Ava Robinson$^5$ \\ Brian A. Smith$^5{^,}{^7}$ \quad Rajan Vaish$^5$ \quad Andrés Monroy-Hernández$^5{^,}{^8}$} 

\def \authors{Samantha Reig, Erica Principe Cruz, Melissa M. Powers, Jennifer He, Timothy Chong, Yu Jiang Tham, Sven Kratz, Ava Robinson, Brian A. Smith, Rajan Vaish, Andrés Monroy-Hernández}

\affiliation{$^1$Carnegie Mellon University, Pittsburgh, USA \quad $^2$New York University, New York, USA \\ $^3$Stanford University, Stanford, USA \quad $^4$University of Washington, Seattle, USA  \quad $^5$Snap Inc., USA \\ \quad $^6$Independent, Mercer Island, USA \quad $^7$Columbia University, New York, USA  \quad $^8$Princeton University, Princeton, USA \country{}}

\renewcommand{\shortauthors}{Reig, Principe Cruz, et al.}

\begin{abstract}
Technology, especially the smartphone, is villainized for taking meaning and time away from in-person interactions and secluding people into ``digital bubbles''. We believe this is not an intrinsic property of digital gadgets, but evidence of a lack of imagination in technology design. Leveraging augmented reality (AR) toward this end allows us to create experiences for multiple people\revis{, their pets,} and their environments. In this work, we explore the design of AR technology that ``piggybacks'' on everyday leisure to foster co-located interactions among close ties \revis{(with other people and pets)}. We designed, developed, and deployed three such AR applications, and evaluated them through a 41-participant \revis{and 19-pet} user study. We gained key insights about the ability of AR to spur and enrich interaction in new channels, the importance of customization, and the challenges of designing for the physical aspects of AR devices (e.g., holding smartphones). These insights guide design implications for the novel research space of co-located AR.
\end{abstract}



\begin{teaserfigure}
 \includegraphics[width=\textwidth]{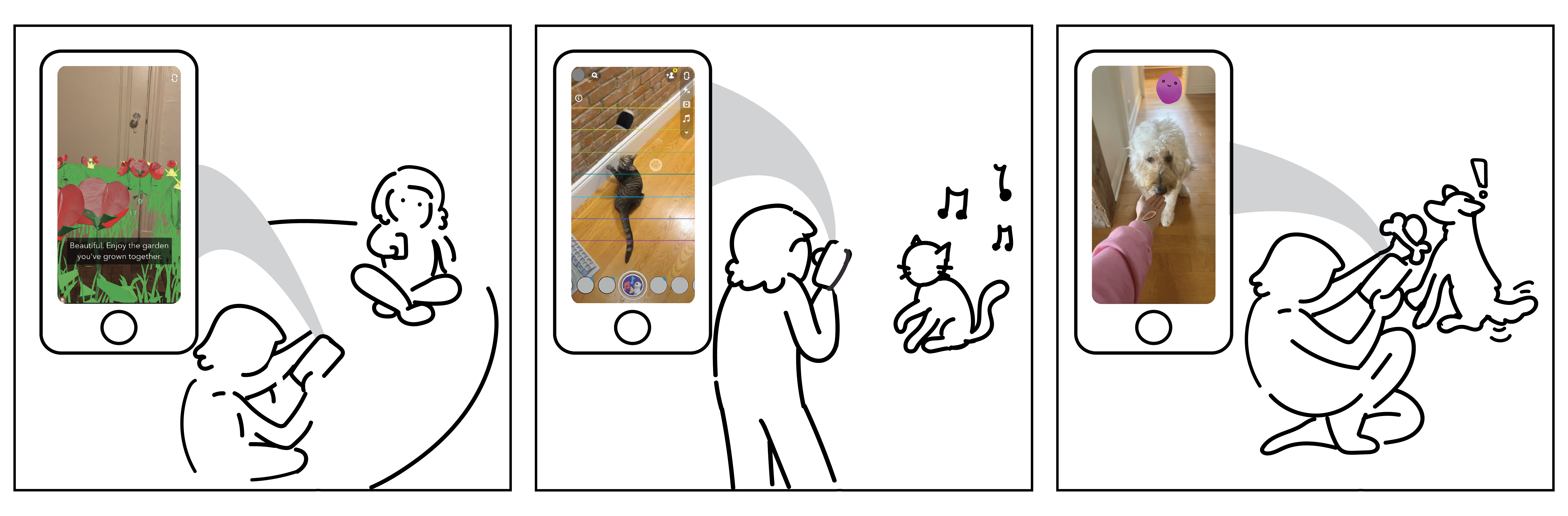}
 \caption{User arrangement and screenshots of our apps: Mindful Garden (left), Compurrsers (middle), and Petpet (right)}
 \Description{A three panel figure showing sketches of user arrangement during play sessions and screenshots of the apps. Left: A sketch showing the user arrangement during Mindful Garden - two users sitting on the ground, facing each other, and both holding their phones. A screenshot showing what the screen looks like playing Mindful Garden - augmented grass and flowers at the end of a breathing session with the text stating ``Beautiful. Enjoy the garden you've grown together``. Middle: A sketch showing the user arrangement during a Petpet session - a user kneeling on the ground facing their dog, holding their phone in one hand, and holding out a dog treat in the other hand. The sketch of the dog shows the dog sitting and attentively facing the user. The screenshot shows the user's screen while playing Petpet through the rear Snapchat camera. The augmented petpet is on top of the dog's head, and an augmented treat is in the user's hand. Right: A sketch showing the user arrangement during a Compurrsers play session - the user is standing and pointing their phone towards their cat. The cat is sitting and has music notes emitting from it. The screenshot shows the user's screen while playing Compurrsers through the rear Snapchat camera with the cat in view and augmented musical strings.}
 \label{fig:teaser}
\end{teaserfigure}

\maketitle

\section{Introduction}

Research suggests that the most enjoyable in-person interactions involve \textit{hanging out} with close ties while engaging in leisure activities such as eating, playing games, making music, and sharing hobbies~\cite{liu2022}. However, digital devices often disrupt these activities. A common view holds that digital technology interferes with co-located moments in ways that reduce their richness or rob them of meaning---as Sherry Turkle put it, ``we'd rather text than talk''~\cite{turkle2017alone}. In this work, we explore how we might design digital technology to \textit{support}, and even \textit{enhance} in-person interactions by \textit{piggybacking} on leisure activities that people already engage in.

Researchers have advocated for ``piggybacking''  when building social computing systems by creating prototypes on top of existing social platforms rather than building something entirely new~\cite{grevet2015piggyback}. We adopt a new approach to piggybacking. We explore  ``piggybacking'' that focuses on the \textit{conceptualization} of the interaction design rather than its implementation. \revis{This involves identifying the core elements and mechanics of familiar activities before augmenting those aspects with novel digital tools.} In this work, we explore what it might mean for technology to piggyback on everyday leisure activities that people engage in when in their personal, private spaces with other \revis{people and/or with their pets}. 

We are guided by the following questions:
\begin{itemize}
    \item How might we foster meaningful, co-located interactions with technology?
    \item What is possible, desired, or valued in technologies designed to piggyback on activities that people already do together?
    \item What are the pitfalls of using technology to piggyback on existing experiences?
\end{itemize}

In this work, we aim to see if we can use technology to enhance people's interactions when they are \textit{physically} together \revis{with humans or pets they are close to}. \textbf{Augmented reality (AR)} is particularly well-suited to this goal: it enables designers to easily extend people's realities in a way that considers their physical worlds~\cite{irl_2022}. AR can take into account its users, their digital environments, and their physical environments to create experiences that are embodied and contextually relevant. While AR has commercially available specialized platforms (e.g., Magic Leap~\cite{magicleap}, HoloLens~\cite{hololens}), it can also be experienced through \textbf{smartphones}. Smartphones are particularly true to the notion of piggybacking because they are already part of many people's co-located leisure time~\cite{liu2022}. Smartphone AR also allows for spontaneity: if a situation arises that lends itself to AR piggybacking, rather than interrupting the flow of the social interaction to set up a technology that might otherwise be stowed in a drawer or powered off, people can simply reach for the devices already in their pockets. In this vision of mobile AR, reaching for a cell phone while hanging out with a friend might signal, \textit{``I have something to add to this moment, and we can experience it together''} just as readily as, \textit{``Time for me to detach from this moment for a scroll break.''} Bringing mobile AR to settings involving spending time together can allow researchers to transform people's ``digital bubbles''~\cite{rogers2014bursting, turkle2017alone}\footnote{\revis{The term ``digital bubbles'' has been used in HCI, social media, and related communities to refer the ``bubbles'' that isolate people from each other when they engage so intently with their own screens that they ignore in-person social context.}} into ``augmented bubbles'' that they can share with others. 

\begin{revision}


This paper introduces three mobile AR apps that we created as design probes. The apps piggyback on three leisure activities: \revis{lull moments with peers} (Mindful Garden), \revis{playing} with cats (Compurrsers), and caring for dogs (Petpet). We created these apps for in-person technology use with social partners (people and pets) with whom people already spend leisure time. Two of our apps are intended for a person to play with a pet; here, we subscribe to the notion that pets are co-located social partners with agency, and experiment with giving them control and agency over technology in their space \revis{ ~\cite{hirskyj2021forming, hirskyjread2016ethics}.} Our exploration of piggybacking allows us to frame the problem of fostering co-located interactions as a form of interdependence~\cite{Bennett2018}, a collective effort between all parties involved---people, pets, the environment itself, and the mobile AR app---to enhance the social bonding occasion. Interdependence and collaboration occur when each party contributes something toward an outcome, and those contributions depend on each other. Our piggybacking apps structure the social bonding occasions on which they piggyback such that each agent involved (multiple human players, pet players, and the app itself) does something different to enhance the occasion.  

\end{revision}

By way of a remote study with 41 participants and 19 pets, the apps we developed helped us uncover insights about the design space of co-located AR designed for piggybacking. The study highlighted the importance of personalized AR experiences and revealed the merits of several design aspects including ``AR coupling,'' i.e., having semantically meaningful relationships between augmented and physical elements, and ``AR mementos,'' i.e., having persistent digital artifacts that people take from the experience. We contribute design guidelines that AR creators and researchers can apply when creating novel AR experiences. Finally, we discuss future research directions for co-located AR. 


\section{Related Work}

\subsection{From ``alone together'' to ``together together'' with technology}
\revis{In her book \textit{Alone Together: Why We Expect More from Technology and Less from Each Other}~\cite{alonetogether2011}, Sherry Turkle famously described how technology, broadly speaking, draws us apart in the physical realm as it draws us closer together in the digital one.} Researchers have described the negative social impacts of ``phubbing'': when someone uses their phone to do a solo activity (scrolling, texting, etc.) in the presence of others~\cite{roberts2016my, chotpitayasunondh2016phubbing, chotpitayasunondh2018effects, ergun2020effects}. This \revise{has been an issue of} debate within the human--computer interaction (HCI) community and beyond it, with studies finding mixed results about the relationship between time spent online and psychological well-being (e.g.,~\cite{chan2015multimodal, tsai2019high, kardefelt2020contextualising}), \revis{designers innovating ways to foster a sense of relatedness over distance~\cite{hassenzahl2012allyouneed}},  and various authors making psychologically and socioeconomically motivated cases against the demonization of smartphones~\cite{balbus_2019, davis_2016}. 

Less attention has been given to how people use smartphones when physically \textit{together}. Past research shows that when co-located, people commonly use one or multiple smartphones and other screens in a ``together together'' way and not just in an ``alone together'' way: they show each other photos and tell stories about the people and events they depict~\cite{lucero2010social, lucero2011pass, van2009collocated}, they brainstorm together~\cite{lucero2010collaborative}, and they co-watch YouTube videos~\cite{sun2017challenges}. In some cultures, it is the historical default for multiple members of a family or social group to share a single mobile phone~\cite{burrell2010evaluating, sambasivan2018privacy}. 

\revise{Scholars within the HCI community have convened workshops \cite{memarovic2012workshop, fischer2016collocated, alharthi2018collaborative, scott2015local} calling for research that focuses explicitly on the role and potential of technology in co-located interactions. In parallel, research has advanced technology that supports co-located social interaction by adding tangible gestural interfaces to existing co-located games to enhance play~\cite{fan2011surprise}, improving gestural interaction during co-located interactions with augmented reality through the use of specific visual cues~\cite{chen2021effect}, facilitating icebreaker activities~\cite{jarusriboonchai2016design}, and fostering rapport among people who are already physically together in \revis{settings ranging from partner intimacy~\cite{vetere2005} to gatherings of} large crowds~\cite{umbelino2019prototeams}.} 
A recent (2019) literature review~\cite{olsson2020technologies} advocated a shift in mentality from designing to \textit{enable} co-located interaction to designing to \textit{enhance} it: rather than existing in tandem with or simply functionally supporting conversation, collaboration, or other forms of face-to-face interaction, technologies designed for co-location can play ``an active role in deliberately attempting to improve its quality, value or extent.'' 

Taken together, these perspectives shed light on a rather optimistic view of the role of tech in our relationships: that introducing technologies to the interactions of close ties can enrich what they are \textit{already gaining} from being physically together. This reveals a need and an opportunity to design expressly for co-location. Work by Dagan et al.~\cite{irl_2022} reveals several design factors that are important for facilitating co-located AR experiences, but does not explore principles for piggybacking on people's \textit{existing} co-located leisure activities and does not advance guidelines for piggybacking itself. In the present work, we explore this new use case for AR to inform the design of technologies that can improve the co-located experiences that people already have.

\subsection{Piggybacking domains}
Per each of our apps' focus, we draw from the literature on animal--computer interaction (ACI) and human--pet--computer interaction (HPCI), virtual pets, and mindfulness in HCI. 

\subsubsection{Cat play and HCI}
Work in ACI and HPCI has explored co-located technology-mediated play between humans and cats~\cite{trindade2015purrfect}, \cite{westerlaken2014felino}. \textit{Purrfect Crime}~\cite{trindade2015purrfect} and \textit{Felino}~\cite{westerlaken2014felino} drew on existing human-cat play to create new playful interactions between cats and their humans. \textit{Purrfect Crime} is designed as a virtual bird-catching competition between a cat and their human using a projector, a Kinect sensor, and a Wii remote. In \textit{Felino}, a human controls the size and movement of sea creatures as they swim by on a tablet for the cat to ``catch''. 
Our AR app \textit{Compurrsers} piggybacks on cats' chaotic behaviors during play. It builds on previous work in that rather than constraining play to a radius, the cat leads the way.   

\subsubsection{Dog play, virtual pets, and HCI}
Past research has also investigated technology-mediated human-dog interaction \cite{wingrave2010early}, \cite{hauser2014improving}. \textit{Canine Amusement and Training} or \textit{CAT} \cite{wingrave2010early} took a serious games approach to teaching humans how to train their dogs using a projector, TV, Wii remote, and sensors. Other work~\cite{hauser2014improving} proposed sensors to support accessible toys to foster play important to guide dog teams' wellbeing. 
\textit{EmotoPet} \cite{wuemotopet} was an environment-sensing AR pet that responded to human speech and gestures. Inspired by these works, our AR app \textit{Petpet} piggybacks on key day-to-day human--canine interactions: giving treats, petting, and calling dogs by their names. 

\subsubsection{Mindfulness and HCI}
For the purposes of this paper, we discuss ``mindfulness'' as a state a person can enter to be more present at a specific moment ~\cite{brown2003benefits}. Research in HCI explored technology-mediated mindfulness for individuals and co-located groups. \textit{JeL}~\cite{stepanova2020jel} used a projector, two VR headsets, and breathing sensors to facilitate mindfulness practices based on breathing exercises. \textit{Inner Garden}~\cite{roo2017inner} used a projector, physical sandbox, physiological sensors, and an optional VR headset to facilitate mindfulness inspired by zen gardening. 
Building on these, our AR app \textit{Mindful Garden} piggybacks on \revis{lull moments with peers by adding a simple and laid-back mindfulness practice to such moments.}

\subsection{Shared digital possessions}
Designing ``piggybacking'' interactions for close ties poses the challenge of durability, or replay value. While replayability is not always necessary for a co-located play experience (escape-room-in-a-box games like \textit{Flashback}~\cite{boardgamegeek} have become popular in recent years despite some of them being playable only once), we believe that short-form playful experiences should be able to be returned to without being entirely redundant. We drew on preexisting notions of persistence in embodied playful experiences  ~\cite{bartle2004designing, broom2019go}, co-creation and interpersonal relationships ~\cite{rouse2020you}, and digital possessions ~\cite{orth2019designing, odom2011teenagers} to incorporate ``takeaways'': each experience delivers as its outcome something that depends on both players' participation and combines aspects of the physical and digital worlds. These ``takeaways'' allow our apps to repeatedly deliver novelty despite their simplicity: they can be returned to again and again and produce different results every time. 


\section{Augmented Reality Apps}

\begin{figure*}
  \includegraphics[width=\textwidth]{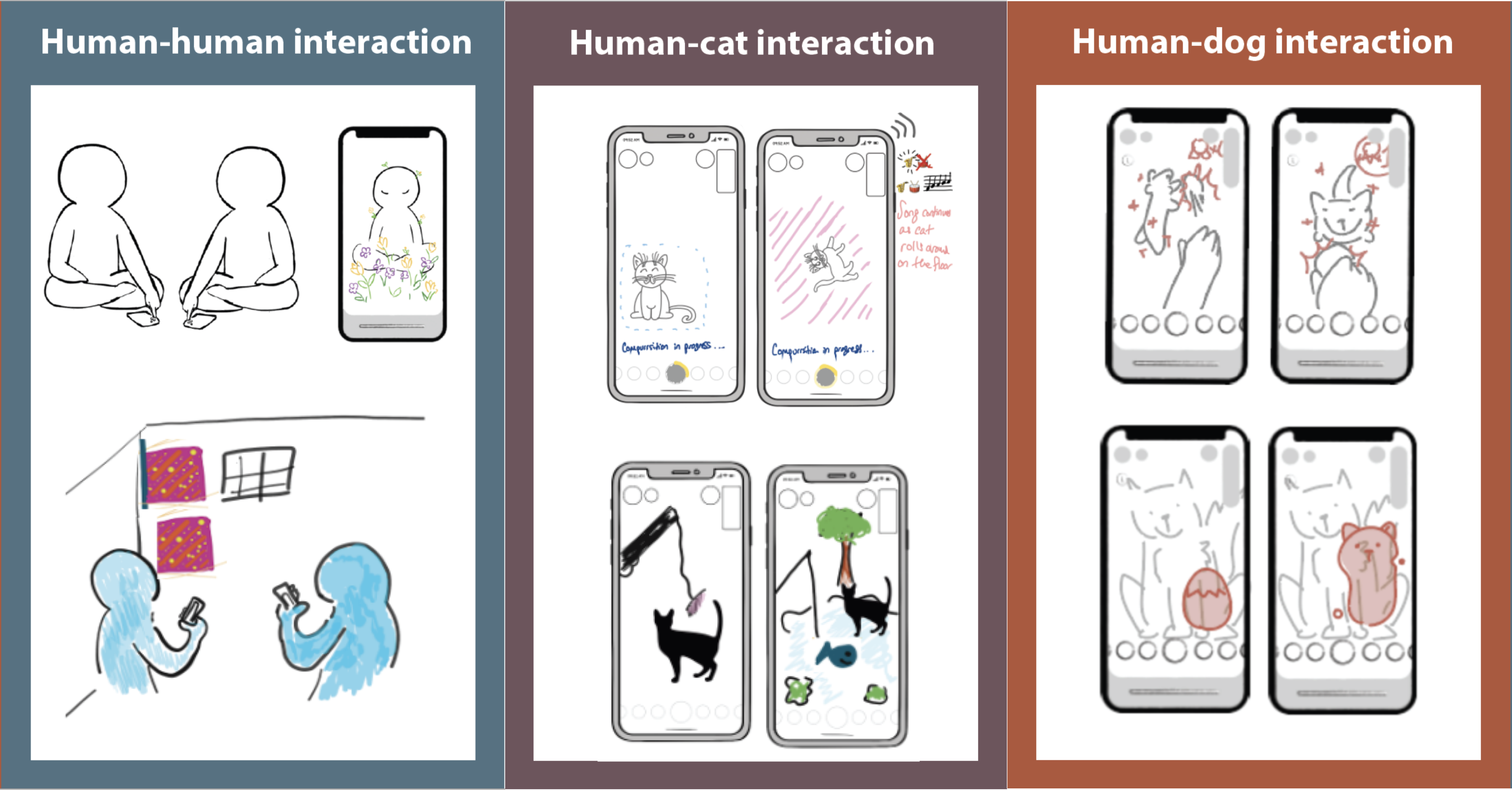}
  \caption{Storyboards showing six of the 45 specific ideas for apps that piggyback on co-located leisure time. Left: two storyboards showing different interactions that involved co-creating something in the environment in a state of calmness while attending to breathing and heart rate; these eventually converged to Mindful Garden. Middle: two storyboards showing a cat moving around and interacting with toys in front of a camera and the app responding with visual and audio augmentations; these eventually converged to Compurrsers. Right: two storyboards showing a person adding emojis to the screen while interacting with a dog and digital animals responding to a person petting and calling a dog; these eventually converged to Petpet.}
  \Description{A three-panel figure showing various diagrams of people interacting with phones and sketches of app screens.  The left panel shows human-human interaction: in the top row, two figures sit side-by-side with their fingers on their camera lenses and an app screen shows a person sitting in an AR garden; in the bottom row, two figures hold their phones up near a wall that has AR squares projected onto it. The middle panel shows human-cat interaction sketches: in the top row, two screens show a cat sitting and lying down sideways in front of a camera while the screen reads ``Compurrsition in progress'' in the bottom row, two screens show a cat toy dangling in front of a cat and an AR pond scene in which the cat appears to be catching fish. The right panel shows human-dog interaction sketches: in the top row, two screens show a hand extending towards a dog and a dog approaching the camera, both with AR shapes emitting from the dog; in the bottom row, two screens show a dog sitting in front of the camera with an AR egg and an AR animal ``between'' the dog and the camera.}
  \label{fig:storyboard}
\end{figure*}

We designed our three AR experiences through a process of ideation and iteration. First, our team identified aspects of multi-person AR experiences that were under-explored in HCI publications and apps on the market. Inspired by these under-explored design concepts, we brainstormed and created storyboards (see Figure~\ref{fig:storyboard}) for 45 ideas for co-located AR apps. As we prioritized the concepts, we abandoned some of our ideas, merged others, and iterated on them as they converged. It was through this process---and especially motivated by Abowd's \textit{identity} (who), \textit{activity} (what), \textit{location} (where), and \textit{time} (when) elements of context~\cite{abowd1999towards}---that we conceived of the goal of \textit{piggybacking leisure time}. Through this process, we also noted the potential of designing AR experiences that are brief in duration and lightweight in requirements, but that persist over time. 



\begin{revisio}
Table \ref{tab:irllenstable} summarizes how each of our apps was designed to support a type of co-located leisure (i.e., spending meaningful time together, or bonding); the specific activity that was piggybacked; and the role of the players and the app in enhancing that co-located leisure activity through piggybacking AR.

\begin{table*}
	\centering
    \small
	\color{black}\begin{tabular}{>{\raggedright}p{0.1\textwidth} >{\raggedright}p{0.12\textwidth} >{\raggedright}p{0.25\textwidth} p{0.45\textwidth}  }

     \textbf{App} & \textbf{Social bonding occasion} & \textbf{Piggybacking activity} & \textbf{Roles of agents in enhancing the occasion} \\

    \hline
    
	Mindful Garden & \revis{Moments of lull/quiet experienced while spending} time with your friend (i.e., ``hanging out'') & \textit{Grow a shared garden:} Practice mindfulness together through a breathing exercise that facilitates the growth of a garden with different flowers representing the players' peaceful/mindful state & \textbf{Each player:} reads the breathing exercise instructions, does the breathing exercise while listening to the other player read, looks around at the resulting garden; \textbf{App:} provides the connection between the phones, guides the meditation, ``grows'' grass and flowers in the players' space \\
	
	\hline
	Compurrsers & Spending time with your cat (i.e., playtime) & \textit{Create a new tune:} Follow the cat with a camera to produce music based on the cat's movements  & \textbf{Human player:} follows the cat around, may encourage the cat to move a certain way; \textbf{Cat player:} moves around the space, may eat or jump or roll around; \textbf{App:} prompts the human to follow the cat, translates the cat's movements into musical sounds  \\
	\hline
	
	Petpet & Spending time with your dog (i.e., playtime and cuddling) & \textit{Grow a digital petpet and keep it alive:} Take care of a digital pet that evolves based on your dog's activity and emotional states & \textbf{Human player:} follows the dog around, calls the dog's name, feeds the dog, plays fetch with the dog; \textbf{Dog player:} eats treats, accepts pets, plays and moves around the space; \textbf{App}: prompts the human to follow/call/cuddle with/feed/play with the dog, ``grows'' the petpet\\
	\hline

	\end{tabular}
	\caption{Summary of each app in terms of the social bonding occasion being piggybacked, the activity through which it piggybacks, and how piggybacking allows each agent (multiple human and pet players and the app itself) to enhance the social bonding occasion.
	}
	\label{tab:irllenstable}
\end{table*}
\end{revisio}

In addition to exploring the notions of \textit{piggybacking} and \textit{persistence}, we also aimed to make meaningful use of co-location and AR. To achieve the former, each app featured an interaction that relied on the physical presence of two players in the same space and would not be the same if done remotely or by just one player. To make meaningful use of AR, the apps incorporated semantically relevant information from the physical world into the virtual aspects of the experience (we call this \textit{coupling}, and explore the impact of it in detail in the Findings and Discussion sections). Our apps were designed as Snapchat ``Lenses'' and built using Lens Studio~\cite{lensstudio}, an authoring tool for creating Lenses (augmented reality filters) for Snapchat users using visual and audio assets, machine learning models, and scripting.

\subsection{Mindful Garden}
\revis{Mindful Garden is intended for moments of lull or quiet experienced while spending time with someone else. It \textit{piggybacks} on these moments by adding to them a short, guided meditation practice consisting of a simple breathing exercise. We chose this activity because mindfulness meditation is a calm interaction with precedent in HCI design (see, e.g.,~\cite{patibanda2017lifetree, dauden2018review, CHITTARO201456}). Through its prompts, the experience also aims to draw attention to an underappreciated aspect of co-location: noticing one another’s physical presence.} On-screen instructions prompt participants to observe each other’s breathing, listen to one another’s voices, and appreciate that another person is physically ``there'' with them. To underscore the \textit{shared} aspect, the \revis{pair collaboratively ``grows''} an AR garden that they can both see. 

The app instructs each player to focus on their breathing, and prompts them to notice their pulse by placing their fingertip over the rear camera lens. This concept is inspired by smartphone apps that use changes in color to ``track heart rate''~\cite{patel_2021}\footnote{Heart rate is not actually tracked in our app; we explained this to participants in our studies.} Covering the lens while growing the garden also prevents players from seeing the full visual of their digital garden atop its physical stage until the mindfulness activity is complete, making for a moment of surprise and delight when the garden reveals itself. 
The interaction requires physical co-presence in that players are asked to read instructions aloud while their partner closes their eyes and relaxes\footnote{Reading aloud can also be done over remote video or audio chat, but feelings of co-presence and the ability to adjust reading pace and volume are not the same in a remote context as in a physical one. Though Mindful Garden could theoretically be done remotely, with the two players seeing the same garden in different spaces, it is designed with the intention for players to share the embodied, face-to-face experience of listening to each other's voices and seeing the same physical space augmented with the same virtual elements.}. The experience is \textit{coupled} as players are asked to sit on the ground while a virtual garden sprouts around them (as if sitting among flowers in a real garden), blossoming in tandem with the players' sense of inner calm. 

Mindful Garden begins by asking the user who opens it (``Player 1'') to choose a friend to play with. Player 1 can then choose someone (``Player 2'') from their Snapchat friends list
Once Player 2 accepts, and the players are connected, the app asks them to look around; when they do, it triggers AR grass to grow on the physical ground. The activity comprises three phases: \textit{Player 2 Breathing}, \textit{Player 1 Breathing}, and \textit{Breathing Together}. First, the app prompts Player 1 to read instructions aloud to Player 2, who is prompted to listen: \textit{``We will use this Lens to share a moment of mindfulness with each other and grow a shared garden in the process. We will each take a turn guiding one another through a breathing exercise \ldots The breathing exercise goes: Breathe in for 4 counts; Hold for 7 counts; Breathe out for 8 counts.''}~\cite{weil2017three}. Player 1 guides Player 2 through the \textit{4-7-8} breathing exercise which repeats four times. Then, they switch, and Player 1 is prompted to close their eyes while Player 2 guides them through the breathing exercise four times. Familiar with the breathing exercise after doing it individually, the two players then do four rounds of \textit{4-7-8} breathing together. At the end of the \textit{breathing together} phase, the app tells players to look around at the garden they've grown. By panning their rear camera around the space, they find their garden now contains three different flower types: One representing Player 1, one representing Player 2, and one representing the two of them together.

\begin{figure*}
  \includegraphics[width=\textwidth]{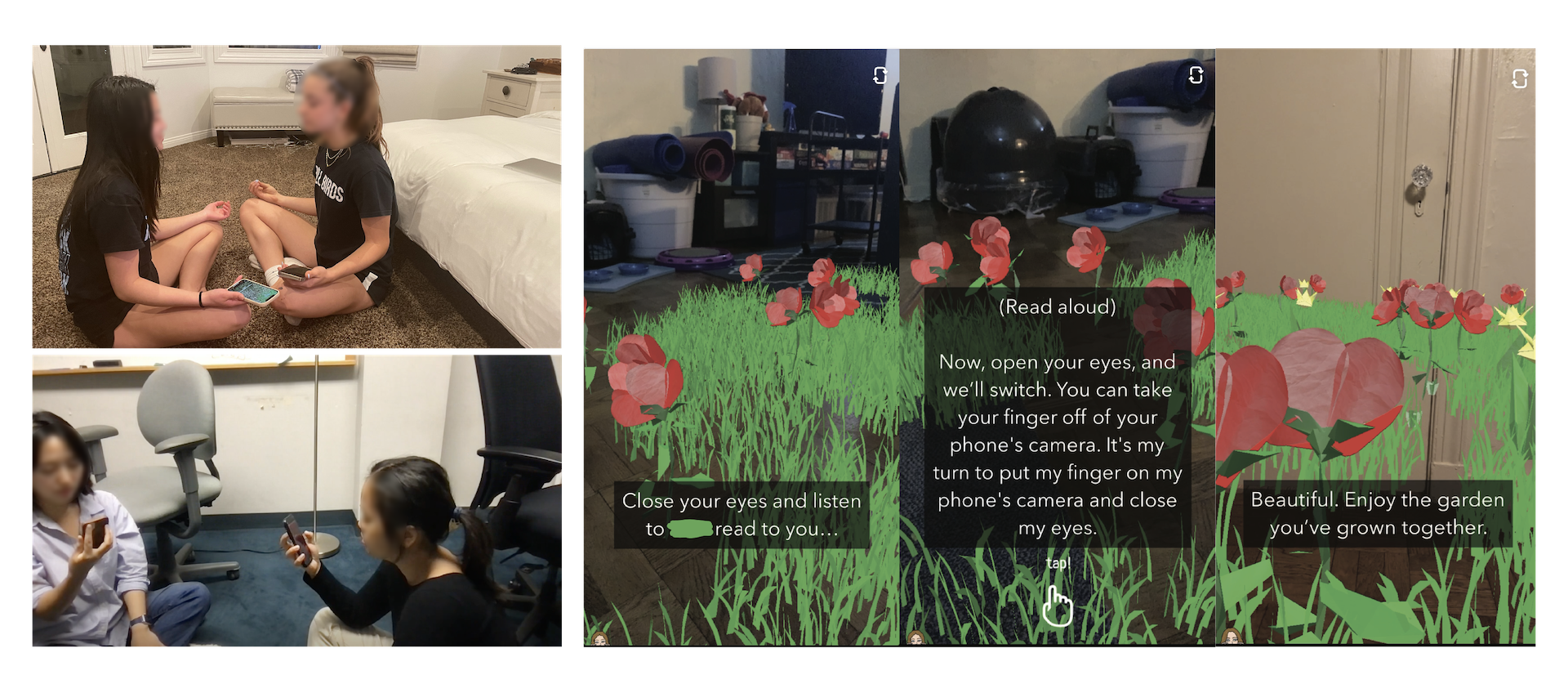}
  \caption{Left: Players using the Mindful Garden app together. Right: Screenshots of the Mindful Garden app instructing players to 1) close their eyes and listen to their partner read to them and 2) read aloud a message to switch roles after going through the breathing exercise three times. The final panel shows the garden that appears after the players complete the breathing exercise (each guiding the other, and then breathing together), containing two types of flowers---red for one player and yellow for the other.}
  \Description{A figure showing two sets of users using the Mindful Garden app together and three screens from the Mindful Garden app during a play session. Left: An outsider's view of two different play sessions. Users are sitting on the ground facing each other and both holding phones using the Mindful Garden app. Right: The background is a photo of a living room taken through the Snapchat rear camera. The augmentation shows grass growing from the apartment floor; there are a few red flowers among the grass. Text on the first screen reads, ``Close your eyes and listen to [name blotted out] read to you''. Text on the second screen reads, ``(Read aloud) Now, open your eyes, and we'll switch. You can take your finger off your phone's camera. It's my turn to put my finger on my phone's camera and close my eyes.'' There is an icon of a hand with the word `tap' below this text. The third screen shows the augmented grass with more red and yellow flowers. The text on the screen reads, ``Beautiful. Enjoy the garden you've grown together.''}
\end{figure*}

\subsection{Compurrsers}

We designed Compurrsers to \textit{piggyback} on playtime with cats, using augmentation to celebrate the eccentricities of their physical behaviors. Cats have a reputation---particularly online \cite{21Tweets5:online, 21WeirdT32:online}---of behaving erratically and acting against their owner's wishes. Compurrsers celebrates, rather than downplays, that well-earned ``personal brand'' with augmentation that responds to the \textit{movement} that occurs in human-cat interaction.  As a cat runs, jumps, and rolls around, its movements trigger music that plays through the app. \revis{If the cat goes outside the frame during play, the text ``looking for cat'' and a scanning animation appear on the screen, and the game resumes once a cat is located.} 

The interaction relies on the physical co-presence of both ``players'' in that the human chooses to play Compurrsers and points the camera, and the cat moves to trigger the music (via the app's cat detection model). The tune's features are determined by the cat's position in the coordinate space on the screen. The audio ``instrument'' used is a cat's pitched meow. Compurrsers is primarily an aural experience, deviating from the typical AR mold of visual in, visual out. Leveraging audio allows a human and pet to experience something interesting together, rather than constraining the human's attention to the phone and not the pet.


\begin{figure*}
  \includegraphics[width=\textwidth]{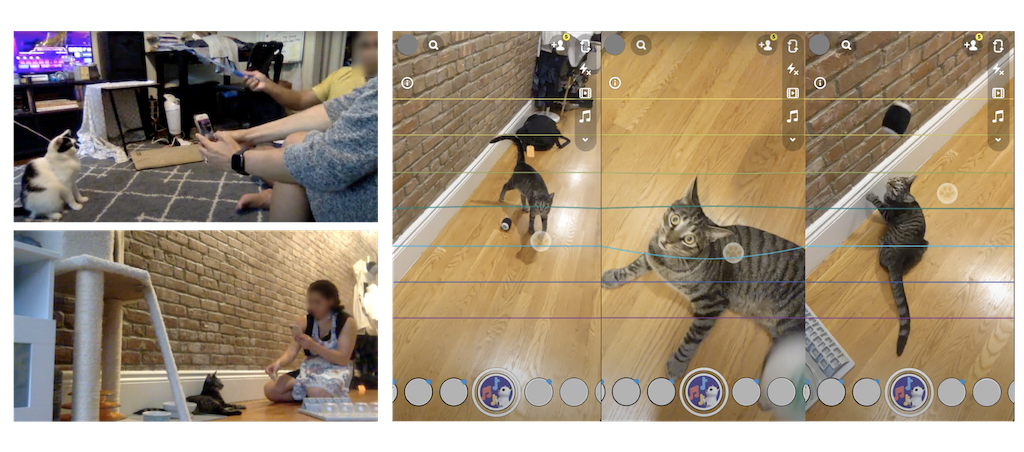}
  \caption{Left: Players using the Compurrsers app with their cats. Right: Screenshots from the Compurrsers app during a play session with the cat triggering music as it moves around.}
  \Description{A figure showing the outside view of two different play sessions of users using the Compurrsers app with their cats and three screens from the Compurrsers app during a play session. Left: An outsider's view of two different play sessions. The top image shows two users sitting on the living room floor and their cat sitting in front of them. One user is holding their phone toward the cat and the other user is holding a cat toy. The bottom image shows a user sitting on the floor, pointing their phone at their cat near a cat tree. Right: The background is a photo of a cat walking down a hallway, looking towards the camera, and playing through the Snapchat rear camera. The augmentation shows musical strings and a cat paw indicating the location of the detected cat. In the second image, one squiggly string indicates that the cat triggered the musical output.}
  \label{fig:compurrsersscreens}
\end{figure*}

\begin{figure*}
  \includegraphics[width=\textwidth]{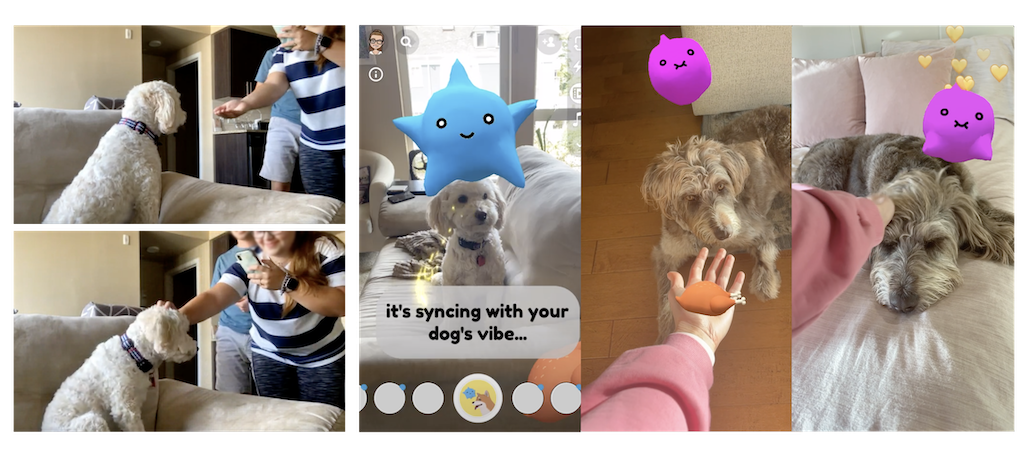}
  \caption{Left: Players using the Petpet app with their dog. Right: Screenshots from the Petpet app during a play session. After the dog is detected, the app says the petpet is ``syncing with your dog's vibe''. In the \textit{feed} interaction, the user extends their hand, and an AR chicken leg is fed to the petpet in time with the dog getting a treat. Then, the petpet responds by emitting gold hearts (far right).}
  \Description{A figure showing the outside view of a user and their dog using the Petpet app and three screens from the Petpet app during play sessions. Left: An outsider’s view of a user interacting with their dog during a play session of Petpet. In both images, the dog sits on a couch and faces the user attentively. The top image shows the user pointing their phone towards the dog and handing their dog a treat. The bottom image shows the user petting the dog while still pointing their phone toward it. Right: Screenshots of the play session from the view of the Snapchat rear camera. The first image shows a dog sitting on a couch being initially detected. The text on the first screen reads, ``it’s syncing with your dog’s vibe…`` and the blue petpet is hovering above the dog's head while sparkles circle the dog, indicating detection. The second screen shows the user holding their hand out to the dog with an AR chicken leg on top of their hand. A purple petpet, in its smallest form with one spike, hovers above the dog. The final screen shows the petpet reacting to the dog being fed with gold hearts emitting out of the petpet and two new spikes being grown.}
  \label{fig:petpetscreens}
\end{figure*}

\subsection{Petpet}

Petpet \textit{piggybacks} on three everyday exchanges between humans and dogs: feeding, petting, and calling the dog by their name. It uses a digital pet or ``petpet'' to accentuate physical displays of affection and attention with a real-life dog. 
Petpet supports AR-mediated human-canine interaction by facilitating persistent play by co-raising a petpet. Both human and dog need to be physically co-present for the app to work. When the user first opens it, they are prompted to find their dog. When the dog is recognized by their phone's rear camera, an egg appears on the dog's head. After a few seconds, the egg hatches and becomes the first form of the petpet---a small, blue, 3D blob with eyes. Over the course of the interaction between the person (or people) and the dog (or dogs), the app uses human hand tracking, dog face tracking, and a dog emotion detection model to determine the next activity. Each interaction triggers a response from the petpet that is intended to mimic what the dog might be doing or feeling. For example, when the dog is fed a treat, the petpet is fed a juicy digital steak, and while the dog enjoys their treat, the petpet emits gold heart emojis while smiling, wiggling, and trilling in delight. \revis{If the dog goes outside the frame during play, the text ``looking for dog'' and a scanning animation appear on the screen, and the game resumes once a dog is located.}

Part of the inspiration for this app was our research team's own memories of Tamagotchi~\cite{tamagotchi}, the Japanese digital pet toy popularized in the late 1990s. Like how Tamagotchi evolved over many play sessions, the petpet evolved as the app detects humans repeatedly responding to its prompts and interacting with their dogs. The petpet changes physical form after the camera recognizes more human-canine interaction: the small blue baby blob becomes larger and grows a spiky appendage but retains the same eyes. In this way, the petpet's development is driven by the dog's emotions and their interactions with their human. At the end of each play session, the app saves the petpet's state so it can be loaded for future nurturing. The human can also take Snaps with their dog and co-raised petpet to share with friends. 


\begin{revisio}
\subsection{Technical Implementation}
We implemented our applications using the Snap AR platform and the desktop authoring environment  Lens Studio. This allowed us to deploy the applications on Android and iOS as Snapchat lenses.  For Mindful Garden, we used Snap AR’s ``Connected Lenses'' functionality~\cite{connectedlenses} to start a shared session for the two players. We used Snap AR's built-in object tracking framework~\cite{objecttrackingdocs} to recognize and track pets~\cite{petdocs}\footnote{Snapchat does not provide performance metrics for animal detection models.}. The pet tracker came with world 3D coordinates for the pet's face. For Compurrsers, we projected these coordinates to the floor plane, and calculated their proximity to each of the ``music note lines'' on the floor (see Figure \ref{fig:compurrsersscreens}). Below a certain threshold distance to the closest line, a note is triggered, or timed out in sync with the backing track's tempo. To test if a note can be played, we prevent repeat notes from playing (so if the cat lingers in the same position, a note will not play continuously). In our tests and during the sessions with participants, the cat detection technology was reliable.
\end{revisio}

For the dog emotion detection functionality, we trained a bespoke convolutional neural network that uses an input image of a dog to estimate the probability of certain dog emotions being present. The possible emotions are ``playful'', ``alert/aroused'', ``appeasing,'' and ``neutral'' (i.e., not exhibiting one of the first three). We arrived at this set of dog emotions after consulting animal behavior experts from the Yale Canine Cognition Center~\cite{doglab}. Dogs exhibiting one of these emotions usually display strong visual cues through body language that can be captured and detected in images. \begin{revisio}To create a ground truth dataset, we hired 30 pet photographers on Upwork.com to take photos of dogs exhibiting each emotion. To better match the real-world application scenario of our model, the photographers were instructed to take pictures using mobile phones. We compiled a dataset of 7,380 images approximately evenly split among the four categories. A team of three expert labelers in our organization created bounding boxes and labels on the images. We trained the labelers with a written labeling guide showing positive and negative examples for each class, and examples of images to exclude from the final labeled set of images. This helped them identify the dog poses that matched the different emotional states. We disposed of the photo if the label did not match the photo category. We used separable convolution with batch normalization as the building blocks of our model (see~\cite{howard2017mobilenets}), reserving 20\% of the images as a validation set. Averaging over five different random assignments of labeled images to the training and validation sets, we achieved precision, recall, and F1 scores of 0.91, 0.69, and 0.79, respectively. The distribution of label counts for each class within the dataset was reasonably balanced, with the label counts per class not differing more than 10\% from each other. Therefore, we did not implement measures to rebalance the data or add specific weights for the labels of each class. Although we did not do a proper evaluation of the quality of our model, in our tests and during the sessions with participants, the dog detection technology was reliable. Unfortunately, the photo dataset is proprietary and, at present, cannot be made available to the public. 
\end{revisio}

\subsection{Participants}
We recruited participants for our remote study through social media and word of mouth. For Mindful Garden, we asked participants to sign up in pairs. For Compurrsers and Petpet respectively, we recruited participants who could bring at least one cat or one dog, and, optionally, another person. We aimed to run between 8 and 12 sessions for each app. \revis{Challenges} with recruitment impacted the number of sessions we ultimately ran: 11 for Mindful Garden (a total of 22 people), 8 for Compurrsers (10 people and 11 cats), and 7 for Petpet (9 people and 8 dogs). This meant that across our 26 study sessions for the three apps, we had 41 human participants and 19 pet participants.

The age range of human participants was 19 to 51 (mean: 27, median: 25)\footnote{These sample population statistics are from 38 of our 41 participants; three did not complete the questionnaire.}. 25 identified as female and 12 as male. Most were undergraduate students, graduate students, or professional engineers. Other professions included designer, registered nurse, nanny, and paralegal. For the pairs of participants (15 in total), 7 were spouses or significant others, 7 were friends and/or roommates, and 1 was a pair of siblings. 12 pairs lived together, and 3 did not. Cats ranged in age from 1 to 11 years (mean: 4) and dogs from 2 to 11 years (mean: 6). The animals were various breeds, and were described by their owners as being varying levels of active, lazy, or both. 

\section{Study Procedure} \label{sec:procedure}
We facilitated participants' use of one of three apps over a Google Meet video call. First, the one or two researchers present\footnote{With two researchers, one led the interview while the other observed and took notes.} introduced themselves and the study. Participants gave informed consent to participate and be recorded, then scanned a code on their own devices to allow them temporary access to the app via Snapchat. They then tested the app for an unspecified amount of time\footnote{The apps differed in levels of structure. This impacted how long they took to play, and how much control users had over the play duration.} and told the researchers when they were done. \begin{revisio}To prioritize cat and dog participants' comfort and agency~\cite{hirskyjread2016ethics}, we conducted tests remotely while pets were in familiar environments with their human(s). We also ended testing when the human perceived that their pet did not want to engage anymore and/or their pet  was no longer interested (e.g. walking away and not returning if called). In cases where this did not happen, we ended testing when the pet participants had played all available interactions of the app or when the human participant(s) felt satisfied with their exploration of it. \end{revisio} We then conducted a semi-structured interview. Finally, we asked participants to complete a short questionnaire collecting demographic information and additional thoughts. Most study sessions took about 60 minutes (a few were shorter). Although these relatively brief exposures did not allow us to eliminate novelty effects, we prioritized using our time to run more sessions, with more apps, in more contexts, which helped us understand how piggybacking AR might work in many different types of settings and among many different types of players. We ran 2-3 pilot sessions for each app within our organization.


In interviews, we asked questions including how participants felt about their co-participant and/or pet during the experience, the value or challenges they found in the physicality of the experience, and what metaphor they would use to describe the role of the physical device in-app. The post-study questionnaire asked questions about interpersonal closeness, with whom else participants could imagine using the app and under what circumstances, and demographic information. The interview guides and questionnaires for each app are included as supplementary material.


\section{Findings}
We analyzed the interview data using a hybrid approach informed by thematic analysis \cite{joffe2012thematic} and affinity diagramming \cite{beyer1999contextual}.  
\revis{We discussed the interviews as they were taking place to begin to interpret them and to determine a natural stopping point for the study given the recruitment challenges mentioned earlier. We then annotated and collectively interpreted the interview transcripts, amassing} several hours' worth of discussion to formulate and debate our understanding of salient themes. We referred to questionnaire data to supplement the interview analysis and confirm our interpretations of what participants said. 

Direct quotes from participants included in this section come from both open-ended questionnaire responses and interviews. We refer to quotes with a letter indicating which app participants used in the study session (\textit{M--}Mindful Garden, \textit{C--}Compurrsers, \textit{P}--Petpet) and a number representing the session ID. Where it is important to include dialogue to provide context for a quote or explain a finding, we use ``A'' and ``B'' to distinguish between the participants. 

Through iterative affinity diagramming and discussion, we identified seven insights about piggybacking co-located leisure activities with AR. \revis{We describe each insight in a subsection below. The findings are summarized in Table~\ref{tab:irlfindingstable} for easy reference.}


\begin{revision}

\end{revision}

\subsection{\revis{AR inspires newfound and renewed motivations to play}} \label{finding:motivation}
First, we found that piggybacking AR provided motivation, awareness, and reminders to engage with each other and with the activities of mindfulness and pet playtime. Participants mentioned that by facilitating interactions that supported familiar activities in novel ways, the apps attracted and held their attention, giving them renewed drive to participate in the activity.

In Mindful Garden, being prompted to meditate together with someone was perceived as especially meaningful. One participant said,  \textit{``It's motivating to have someone at your side when you're trying to do better for yourself''} (M13). Another talked about how co-location could improve focus: \textit{``I think it's good that we're doing it together because I feel like sometimes if you're doing it alone, then maybe you can't really concentrate on that''} (M12). Finally, positive feedback from the virtual flowers could encourage replays: \textit{``Normally, when you do breathing exercises, you don't see something happen and you just feel good, but now you see a flower bloom, that motivates you to do it every day''} (M13).

In Compurrsers, the promise of new and different tunes motivated owners to keep playing with the app, which also meant attending more to their cat. One participant described trying to make sense of and control the tune that was generated: \textit{``The music that was in the background, I liked that a lot, it was upbeat. It did motivate me to try and play a little bit''} (C2) (this participant used ``play'' as in ``play an instrument'' rather than as in ``play the game''). AR could also  enliven play sessions at the point when the owner has become bored or tired but the cat has not. One person explained: \emph{``I will get bored and then he will be like, `Oh, yes, keep playing, keep playing,'' but if I can use the [app] as a way to entertain myself as well while playing with him, maybe he will encourage me to play with him longer''} (C8).  In Petpet, \revise{people mentioned that} the need to maintain the petpet's health and help it evolve---something only possible with the active participation of both human and physical dog---could keep \revise{them} coming back to the app, and, more importantly, \revise{back to} their dog. As one participant put it, \textit{``This is a great tool for getting people to be aware like, `Oh shit, I have a pet, I should play with my pet,' and then put the phone down and play with your pet''} (P12). \revis{The pet apps also encouraged novel kinds of group HPCI. Several Petpet sessions had multiple human participants to begin with; in these sessions, individuals reacted not only to the petpet, but also to each other's reactions to the petpet, which often resulted in laughter and banter. In a handful of cases, multiple members of a household (e.g., partners, roommates) signed up for Compurrsers and Petpet sessions. In these cases, people sometimes passed cell phones back and forth during the session and joined ongoing interactions (e.g., treat-giving) partway through, making the experience more social and more lively for all involved.} 


\subsection{\revis{Participants had platform-driven expectations about what mobile AR experiences entail}} \label{finding:playtowin}
We noticed that participants brought different ``baggage'' that reflected preconceived notions about what phones are meant to be used for. Many expected the apps to be gamified or competitive, and some were thrown off by the lack of emphasis on ``snappable'' moments. This was most prevalent in Mindful Garden---possibly because it involved two players. Several participants (M7, M10, M12, M14) mused about adding a competitive element to the garden. In one case, this led to a debate about whether this was or was not in the spirit of the app: \emph{``I guess I was trying to go and see whether we could see whose flowers were more at the end of the day.''} (M10A); \emph{``Then it becomes competitive. This is a mindful game, though. It's not about the number of flowers. It's just about growing flowers''} (M10B).
The use of Snapchat as a platform also affected expectations about what the apps were ``for''. One participant who was familiar with Snapchat Lenses asked about Mindful Garden, \textit{``But why is it a Lens? It didn't want anything from me? I would never be like, let's open Snapchat and open a Lens to help us meditate''} (M8). 
These assumptions and preferences about the ``game'' being ``winnable'' raise the question of how  piggybacking AR can be designed to get away from the notion of ``mobile app means gaming means trying to win.'' 

\begin{revision}

\end{revision}

\subsection{\revis{Augmentation illuminates information that is otherwise hidden}}
\label{finding:newchannels}
\revis{Participants commented that the apps allowed them to connect in new ways, using information about one another that would be either unavailable or unattended to without the augmentations.}

\revis{In Mindful Garden, participants appreciated that they could interact with a familiar person, in a known space, in a new way. The app brought a novel sensory experience to a familiar activity: the sight, sound, and touch sensations of sitting near someone and sharing a mood and mindset were embellished with a new visual that embodied their sense of focus, calm, and co-presence. In M8, a participant said: \textit{``[It was] like we were giving each other a little gift of mindfulness.''}} \revis{In Compurrsers, participants liked that} the app gave the cat the reins: \textit{``Pets only have so many ways to communicate with you, so if the app can be used as a way for my cat to communicate with me or present some new human-cat interaction, I'm definitely down for the cat to take the lead''} (C9). 
\revis{Cats sometimes responded negatively when they heard the musical meows---in C10, a cat was terrified of the sounds and ran away from her owner each time she tried to restart Compurrsers.\footnote{In this session, the owner tried various ways to engage the cat, such as giving treats and moving to a different room. When the cat continued to react negatively, the owner stopped trying to play the game; we then discussed the brief experience they had had with Compurrsers and why the participant believed it had been unpleasant for her cat.}} In Petpet, participants \revis{commented} that the petpet gave them new reasons to give their dog attention and affection. They found joy in the prompts to both focus on their phone and check on their dog: \textit{``Trying to get your dog's vibe was just really cool and it really sucked me in because I was like, `What \textup{is} my dog's vibe?' [chuckles] Feed time sounds about right''} (P6). When the dog emotion detector consistently read ``appeasing'', the human felt that the ``vibe sync'' was accurate for the dog's mood, and enjoyed tuning into and acting on the desire for treats via the petpet. 

\revis{Participants appreciated that both pet apps} added new channels for engagement and communication without asking too much of the players. They enjoyed that cats and dogs could exist in their natural habitats---playing in their usual ways and not themselves needing to acknowledge the technology---while people could see, hear, or feel something new, brought about by the \revis{technology. For example,} C6 said, \emph{``We never get to play with our cats and our phones \ldots it's really fun to mix the technology with the \revis{pets \ldots since} they're not looking at the phone it was a lot more intuitive for them.''} (C6). 
\subsection{\revis{Holding a cell phone impacts the user experience of piggybacking}}
\label{finding:physicaldevices}
Our apps faced the dual challenge of \revis{justifying the presence of mobile devices in existing interactions and then }deploying them in a way that did not detract from the intended experience. We found mixed responses as to whether the physical setup and use of the phone was a help or a hindrance in terms of appreciating the piggybacked occasion. 

In Mindful Garden, participants were prompted by the on-screen instructions to focus on their breathing, listen to one another's voices, and meditate together. However, some participants ended up more focused on the physicality of interacting with their phones. Some were distracted by the physical discomfort or awkwardness of holding it. 
\revis{Others} did not want to touch their camera lens \revis{(\textit{``I don't like leaving fingerprints''} (M5)), or} found that their choice of phone case made touching it difficult. 
The requirement of \textit{two} phones also had pros and cons. \revis{In M10, participants compared their gardens when they looked at each other's screens, and one person said that their partner's was \textit{``so much prettier''}. They discussed how the temptation to compare the two screens} made it harder to buy into the promise of a moment of mindfulness that was distinctly theirs. \revis{In M6, however,} a participant stated that they liked being able to compare gardens, even if they ended up different. They appreciated the novelty: \textit{``I never really look at someone else's phone as much''}. 

\revis{In Compurrsers and Petpet, participants reflected on the differences between ``natural'' (quote marks are our own) human-pet-phone configurations that they typically find themselves in (e.g., photographing a cat while it is asleep, taking selfies with a dog) and the ``unnatural'' human-pet-phone configurations required by the apps. C10 hypothesized that her cat did not like when her cell phone physically came between them: \textit{``she’s seeing this big black box, and it’s making scary noises \ldots it became like the thing she wanted to avoid.''}} P9 said they'd like to be able to take a selfie with the dog and the petpet together, and P9 felt that because the app used the rear-facing camera---instead of the front-facing camera, which they use to take selfies with their dog, which requires getting physically close---the dog wasn't involved enough. In these cases, drawing attention to a dog's physicality was not satisfying enough; people are used to having their dogs physically near when interacting with them and wanted that to be the case for this app too.

Across all three apps, participants had more difficulty paying attention to the physical presence of their partner when a design aspect or technical glitch reminded them that their phones were also, in some sense, a physical, co-located party. Successful connections were made when participants saw partners \textit{through} the phone screen rather than seeing them \textit{on} the phone screen.

\subsection{\revis{Participants wanted AR to not just augment moments but make them persist}} \label{finding:mementos}
\revis{In all three apps, participants talked about continuity across multiple plays. They also desired ways to save and download mementos---something by which to remember a piggybacked moment in time.}

\revis{Mindful Garden's participants suggested that being able to return to the same garden would make the experience substantially more meaningful. For example, one participant (M8) said: \textit{``I would like it if it wasn't just a one-time thing because \ldots I feel like everyone could get into it. Like hey, the more people participate, the better, bigger your garden grows, and then over time, people can visit it.''}}
\revis{Participants} were especially focused on the notion of downloading, saving, and sharing something after trying Compurrsers. One participant talked about how the music from the app could be an addition to the videos they already share with their friends of their cat doing different things: \textit{``[I'd] probably send it directly to people if I wanted to crack them up or share, `This is what [my cat] was up to today'''} (C10). Another reflected on how the combination of an interaction designed specifically for a person and a pet, a persistent artifact, and the fact that these were designed for a cell phone---something they already have and use---made the app feel like something special:  \textit{``Now it feels like I could actually include my pets in like doing tech stuff, things that have to do with technology and I got to save the videos \ldots If it's free and then you get to save it onto your phone, it's just like you're doing something like virtual reality or AR and you don't have to go searching for something like this''} (C8). 
\revis{In Mindful Garden and Petpet, participants talked about wanting to take screenshots and videos of flowers growing and dogs and petpets receiving treats and cuddles to look back at later or share with friends.} 



\subsection{\revis{Tightly-coupled digital and physical signals bring about discovery and enjoyment}}
\label{finding:coupling}
Our apps explored a few different approaches to tightly-coupled semantic relationships between physical and virtual aspects of AR experiences.  We found that people appreciated when they could easily connect and interact with connections that the apps made between the physical and digital worlds. For one participant (C2), figuring out the connection between the cat's movements and the resulting sounds was part of what made Compurrsers feel engaging and fun: \textit{``I'm like, `I want to do something different now. I want to experiment. Oh, this noise happened. Why did it happen?'''} One participant, P11, was entertained by the way the petpet's state mirrored their dog's: \textit{``I think the reactions were pretty good, because, I feel like sometimes when [my dog] wanted another treat and I didn't give it to him right away, it had those exclamation marks saying like, `Feed me, feed me.' [laughs] I think they really made sense. It mirrored what [my dog] wanted or liked, or did not like.''} Another, P9, enjoyed ``troubleshooting'' the input-output relationships in Petpet together with their dog:\textit{ ``I don't usually try to sit down and figure something out while when my dog is trying to play with me, that was nice''} (P10).

\revis{On the other hand, when} participants did not understand the relationship between physical triggers and augmentations, they did not grasp or appreciate the piggybacking as much (or at all). This was most obvious with Mindful Garden, where participants searched for a relationship that they could not find between what they were doing and the growth of the flowers. One participant of M7 explained that even though \textit{``it was nice to see a tangible outcome''}, the garden was lacking \textit{``an intuitive connection with how much the flowers would blossom.''} \revis{Participants also commented on forms of temporal and spatial coupling: they wanted the garden to grow gradually rather than suddenly, and they wanted to link the garden to a physical place via GPS.} 
In Compurrsers, some participants said that their inability to meaningfully connect the sounds to cat-generated inputs was a negative aspect of the \revis{experience. 
Some} Petpet participants could not readily identify the reasons for the petpet's responses to the dog's movements. This made the experience feel more like a puzzle \revis{for a human to solve alone} than a piggybacking embellishment on leisure time \revis{for humans and dogs. 
For example,} P8 \textit{``felt like the petpet mimicked my dog's emotions and stuff when I scanned it, but then after that, it felt like just another pet that I was taking care of. I feel like when I was interacting with the petpet, I didn't really interact with my dog as much.''}

\begin{revision}

%

\end{revision}

\begin{table} 
        \centering
        \color{black}
        
	\caption{\revis{Summary of our findings, which correspond to Sections~\ref{finding:motivation}--\ref{finding:customization}.}}
	\begin{tabular}{l p{0.8\columnwidth}}
    \toprule
	\textbf{Label} & \textbf{Finding}  \\
    \midrule
            
	F1 & AR inspires newfound and renewed motivation to play \\
        \hline
        
        F2 & Participants had platform-driven expectations about what mobile AR experiences entail \\
        \hline

        F3 & Augmentation illuminates information that is otherwise hidden \\
        \hline

        F4 & Holding a cell phone impacts the user experience of piggybacking \\
        \hline

        F5 & Participants wanted AR to not just augment moments but make them persist (through continuity and mementos) \\
        \hline

        F6 & Tightly-coupled digital and physical signals bring about discovery and enjoyment \\
        \hline

        F7 & Participants wanted personalization linked to identity and preferences \\
        \bottomrule

	\end{tabular}
	\label{tab:irlfindingstable}
\end{table}

\subsection{\revis{Participants wanted personalization linked to identity and preferences}} \label{finding:customization}
\revis{Participants} thought that piggybacking AR should allow people to choose settings and avatars that represent their personalities, relationships, and preferences. \revis{This could both enrich individual app plays and make multiple plays feel like a continuous, personal experience.}  

\revis{In Mindful Garden, participants discussed how they might be drawn to the garden by the promise of special flowers for special points in time, like an \textit{``anniversary flower''} or \textit{``flowers for each year''} (M6). Some wanted to tie the garden to a physical location in space using GPS or object landmarks (as mentioned in Section ~\ref{finding:coupling}) so that it could only be returned to by the two people who created it when they were together in that space. Participants also discussed how} being able to customize the colors and types of flowers in \revis{the garden} would make it feel personal and more like the product of a shared experience between them. M10 said that they would have liked to see \textit{``a statue or something, a bird that we both see---then we could know for a fact \ldots that it's the same garden''}. 
\revis{In} Compurrsers, many comments about personalization and customization centered on the possibility of the music reflecting the human's musical creativity or preferences.  C5 said, \textit{``if there were a different instrument like tones or more like precise movements you can make, then feel like I could actually really enjoy something like this.''} \revis{C9 said: \textit{``As someone who just revels in chaos, there's not enough technologies out there that just produce chaos so I'm always excited to see that.''} C9 also} suggested that the music could reflect the cat's mood or personality: \textit{``Since my cats are being so lazy right now, it'd be nice to have a low-key waltz [chuckles] or meditative music to match their lack of activity.''} 
In Petpet, participants commented that the more the petpet's behaviors reflected characteristics of the dog (its appearance, age, personality, emotions, or needs at a particular time), the more significant the relationship between the dog and the petpet would be. For example, one participant said, \textit{``Not all dogs look the same. If it shows up looking like a golden retriever and I'm playing with my completely black, total mutt of a dog, it's not going to look the same,''} (P7), and another said, \textit{``[If] every time he even had like a tiny piece of kibble he would lick his lips afterward, it'd be cute''} (P10).


In sum, people want a signal that the app is doing something for the specific people and pets who experience it over repeated interactions. In broad terms, HCI touts the value of having easily-discoverable options beyond the default as a way of designing for individual users' abilities, preferences, and patterns of use. In AR applications, customization can serve this purpose, but it can also do more: it can embody the notion of \textit{``This reflects something that I am doing, here, with you, now''} in the design, which is central to the experience that ``piggybacking'' seeks to provide.

\begin{table}
        \centering
        \color{black}
        
	\caption{\revis{Summary of our design implications for supporting piggybacked co-located leisure via AR, along with the finding(s) from which each is derived. The findings are summarized in Table~\ref{tab:irlfindingstable} and the design implications are elaborated in the Discussion.}}
	\begin{tabular}{p{0.7\columnwidth} r}
    \toprule

	\textbf{Design implication} & \textbf{Finding(s)}  \\
        \midrule
            
	Calibrate expectations and mood by setting the tone within the app or experience. &  F1, F2 \\
        \hline
        
        Find meaningful and delightful ways to piggyback physical inputs by semantically connecting them to digital outputs. &  F1, F6 \\
        \hline

        Emphasize physical co-presence and human and pet embodiment. & F3, F4 \\
        \hline

        Afford personalization, persistence, and memento-making. & F1, F5, F7 \\
        \hline

        Avoid blocking natural human-pet spatial configurations and interactions. & F2, F4 \\
        \hline

        Be wary of device and design elements that may cause physical discomfort. & F4 \\
        \bottomrule

	\end{tabular}
	\label{tab:irlimplicationstable}
\end{table}

\section{Discussion}

Our study revealed several insights about how AR technology can enhance rather than detract from everyday leisure time that people share.  In this section, we interpret \revis{those insights} to pose design implications for creating AR experiences that piggyback on close ties' existing interactions. \revis{We structure these around our three research questions. The design implications and the findings from which they are derived are summarized in Table~\ref{tab:irlimplicationstable}.} 

\begin{revisio}

\subsection{How might we foster meaningful, co-located interactions with technology?}

\subsubsection{Calibrate expectations and mood by setting the tone within the app or experience}
In our study, introductions were useful for ``getting into the zone'', or the mood on which the apps aimed to piggyback. In Mindful Garden, the first player to read aloud needed time to process the instructions and get used to verbally leading their partner, and both players needed time to get into the headspace of quietness and mindfulness. In Compurrsers and Petpet, the lack of an introduction sometimes broke flow: participants had to figure out what the app expected of them and their pets, and then had to get situated to deliver that, which sometimes meant rousing their cats from sleep or getting treats and toys. 
People sometimes had miscalibrated expectations about what the app would ask (Section~\ref{finding:playtowin}), and enjoyed the experiences more once they had adjusted their expectations to fit the app's intended ``vibe''. Therefore, we recommend that designers of piggybacked experiences incorporate a ``burn-in'' phase: an interaction that helps the users understand the app's intent and sets a tone for the experience to come.
\end{revisio}

\subsubsection{\revis{Find meaningful and delightful ways to piggyback physical inputs by semantically connecting them to digital outputs}}

Co-located leisure activities are multi-modal experiences. They leverage a variety of our emotions and senses including vision, hearing, touch, and smell. In this paper, we explored multiple ways of mapping physical inputs (changes in the observable physical world) to AR outputs. Petpet, for example, leveraged pets' emotions as an input, amplifying this unique aspect of users' physical experience and visualizing that aspect via a digital pet. We found that participants strongly resonated with how extensively the apps integrated their physical context into the augmented experience (Section~\ref{finding:coupling}). Participants were excited about Compurrsers' auditory focus, which allowed their cats to experience it as well (they heard and reacted to the sounds) (Section~\ref{finding:newchannels}) and challenged their existing assumptions about what AR apps are. We recommend that piggybacking AR apps make deliberate use of coupling in mapping physical-world inputs to augmented-world outputs in multiple modalities. Previous work in AR has leveraged semantic associations between virtual and physical elements in a user's work environment~\cite{semanticadapt}.  Work in VR interaction design from the disability community such as Canetroller~\cite{canetrol34:online} and Acoustic Minimaps~\cite{AcousticMinimap:online} has made strides in augmenting multiple modalities. Future piggybacking AR apps should continue to go beyond the visual, and could even draw on unnoticed or ``invisible'' inputs such as emotions or biosignals~\cite{liu2017supporting} that are already present in users' leisure activities.


\subsection{\revis{What is possible, desired, or valued in technologies designed to piggyback on activities that people already do together?}}

\subsubsection{\revis{An emphasis on physical co-presence and human 
and pet embodiment.}}
Where our apps succeeded in drawing out physicality, participants expressed joy and understanding; where they did not, participants expressed confusion and felt that phones were barriers to rather than facilitators of in-person interactions. For example, participants enjoyed when they attended to one another's voices and breathing patterns when growing their gardens (Mindful Garden). \revis{Future versions of Mindful Garden could potentially detect lulls or slow moments between friends---opportunities to appreciate each other's presence, without the distractions of other topics or activities---and insert its augmentations into that moment, turning the mundane and ignored into something that can be noticed and valued. Participants also appreciated the emphasis on physical human-pet interaction, expressing delight when} they were encouraged to touch their dogs to help the petpet evolve (Petpet) (Section~\ref{finding:newchannels}). We also identified obstacles to drawing out physicality: preconceived notions about what to expect when opening a Snapchat Lens (Section~\ref{finding:playtowin}) and instances of weak coupling between digital outputs and physical inputs (Section ~\ref{finding:coupling}) made it harder for people to notice or appreciate how the apps highlighted physical co-presence. We therefore posit that a mobile AR app's success or failure in piggybacking physicality hinges on whether the phone directs attention toward physical co-location or away from it (Section~\ref{finding:physicaldevices}). This tracks with related work in digital game design: in Frolic, a mobile game that encouraged young girls to engage in physical play, the phone functioned as a facilitator so that it would not be a distraction~\cite{Frolic:online}. Future work should examine how design elements of AR apps and AR-enabling devices can highlight or undermine physicality. Here, there is an opportunity to draw on research that examines how VR experiences can foster a sense of embodiment in technology-mediated environments~\cite{embodiment:online}.

\subsubsection{\revis{Affordances for personalization, persistence, and memento-making.}}
Participants felt that AR had to piggyback on the interaction between \textit{them, then and there}---and not just \textit{anyone, anywhere}---to avoid being noise and instead be special. They also wanted more ways to control the outputs of the apps both during play (Section ~\ref{finding:customization}) and after (Section ~\ref{finding:mementos}). \revis{Our findings about preservation through self-expression and preservation through mementos are interrelated: by giving users something personal to save, apps can create personal and even sentimental souvenirs rather than generic participation trophies.} We therefore recommend that piggybacking AR apps give participants personalization options and mementos that last beyond the end of the immediate experience. This aligns with past work about how physical artifacts generated during computer-mediated play are valued after play~\cite{Loominar90:online} and how digital artifacts from gameplay experiences can have lasting value~\cite{KindWord16:online}. In future iterations, Mindful Garden could let participants choose their own flowers and activate only in specific locations, Compurrsers could allow people to pick instruments and background music that reflect the cat's mood, and Petpet could offer petpet avatars that resemble the live dogs. Personalization is especially important for piggybacking; aspects of how people engage in leisure activities \textit{without} AR can (and, in fact, need to) make their way \textit{into} AR to make the piggybacking feel genuine. For example, if a person usually meditates near a cactus plant, then their garden could have cacti instead of tulips. If a person usually plays the ukulele and films their cat's responses, their Compurrsition could sound like a ukulele. 

\subsection{\revis{What are the pitfalls of using technology to piggyback on existing experiences?}}

\subsubsection{\revis{Blocking natural human-pet spatial configurations and interactions}}
In Compurrsers, participants valued how easily they could integrate the phone into the piggybacked interaction. They went from simply watching their cats (and occasionally petting or throwing something to them) to \revise{co-creating music while} watching their cats---and the phone enabled this without getting in \revis{the} way too much. In contrast, in Petpet, it was more difficult to manage holding phones, physically interacting with dogs, and monitoring the dogs' responses all at once (Section~\ref{finding:physicaldevices}). The design implication we take from this is: technologies that piggyback on playtime with pets should be used in ways that do not interfere with humans' ability to look at and touch their pets. Designers of piggybacked experiences should start with scenarios that already involve a phone and a pet (such as taking selfies, filming a cat zooming around, or trying to create viral videos) and build off of \revis{those to create novel} interactions that work with pets' existing behaviors. Forcing a device into an interaction in which the device is usually not involved at all will likely detract from piggybacking. 
Prior HPCI innovations helped humans and animals find joy when they could interact with the technology together (vs. just the human using the device)~\cite{trindade2015purrfect, westerlaken2014felino}. Our design recommendation adds to this the idea that HPCI can work well when it allows pets to do what they already do while also giving humans something new to appreciate without monopolizing the humans' attention. We suggest that future work continue to explore how mobile AR technology might unobtrusively facilitate cross-species play at a low cost and with minimal setup. 

\subsubsection{\revis{Causing physical discomfort}}
Participants were uncomfortable using Compurrsers and Petpet for more than several seconds at a time due to these apps' requirement that phones be continuously held up (Section~\ref{finding:physicaldevices}). Different people who use piggybacking AR apps will have different body types and tolerances for the physical movements that apps involve. We draw from this the following design recommendation: keep piggybacking mobile AR apps lightweight and short to make them more accessible and comfortable for more people.
Physical discomfort also arose when players were asked to do unnatural things with their phones, such as when they placed their fingers on the camera lens and dirtied the Lens when playing Mindful Garden. Whenever a piggybacking augmentation requires hardware components to serve unusual purposes or be put in unusual positions---e.g., players placing their phones face-down on the floor or touching their camera lenses---it is especially important for designers to account for physical comfort. One way to do this is to make these ``unnatural'' interactions part of the app's narrative. As a retrospective example, Mindful Garden may have been more effective if the experience concluded with screens that instructed participants to wipe down the camera lens and stand up and stretch. Likewise, Compurrsers could have limited the length of each Compurrsition to a certain number of seconds. In general, AR designers should choose design metaphors that allow the AR device to set the stage and then get out of the way, making itself truly ``visible'' (noticeable) again only to provide structure and guidance when needed. Work in AR/VR has considered how to incorporate natural, expressive gestures for AR/VR interactions~\cite{youtubeHandInterfaces} and examined the opportunistic design of tangible AR interfaces~\cite{opportunistic}. Some AR technologies that use headsets, such as ARcall~\cite{ARcall:online} and Friendscope~\cite{Friendscope:online}, have leveraged this approach to provide novel experiences without asking users to perform unusual physical actions. We emphasize that designers of mobile AR apps need to account for this too, even though the platform they are working with appears commonplace and convenient to hold and use.

\subsection{Limitations}
Our Lenses had some technical drawbacks: the flowers in Mindful Garden did not always perfectly align with the floor, and the dog emotion detector only had four classes. We do not believe these detracted from participants' experiences aside from being minor inconveniences. In our study, each of 41 participants briefly played with one of our apps. As discussed in Section \ref{sec:procedure}, this approach allowed us to reach more participants than we would have been able to with longer or repeated sessions, and we believe it resulted in novel and valuable insights for piggybacking AR. We acknowledge, however, that giving participants only a single encounter with the apps does not give us a sense of how people would engage with them over time. 
Future work could expand on this research with longitudinal studies focused on understanding the influence and implications of piggybacking AR over time. 
While we were able to solicit feedback on our apps from people of many different ages and backgrounds in a variety of physical settings, our sample size was still relatively small, and our participants self-selecting; our insights may not reflect more diverse perspectives. Our interpretations of what participants said about our apps and what this means for piggybacking AR are also limited by our own perspectives. Finally, we only probed three co-located interaction scenarios, all of which were social in nature. There are other categories of meaningful co-located interaction---e.g., professional, educational, or health-related settings---that we did not address. Future work could investigate piggybacking AR in these other contexts.
 

\section{Conclusion}
\revis{In this research, we explored the potential for augmented reality technologies to contribute to co-located leisure time by ``piggybacking'' on naturally occurring scenarios among people and pets. We defined \textit{piggybacking} as a way for social AR to embellish existing interactions by adding something to what people and pets are already doing rather than asking them to do something entirely new. We developed three AR applications that ``piggyback'' on activities that people already do with their close ties. The first, Mindful Garden, piggybacks on sharing quiet moments with peers (and, to an extent, practicing mindfulness). The second, Compurrsers, piggybacks on playing with cats, and highlights their ``brand'' of independence and chaos. The third, Petpet, piggybacks on playtime with dogs, and emphasizes attentive human-dog interaction. We shared the apps with 41 people and 19 pets with the goal of understanding what it takes for an AR technology to genuinely piggyback an existing co-located situation (rather than introducing a new activity). Through observations and interviews, we identified several new insights relating to how technology can piggyback on commonplace interactions in engaging and meaningful ways. Based on these insights, we pose six design implications for creating piggybacking AR technologies. Our work contributes} novel findings about co-located multi-user AR, three new apps designed to support meaningful AR-mediated interactions among people and pets, and several design guidelines for piggybacking AR. 



\begin{acks}

We thank all of our participants for their time and feedback. We are grateful to Jack Tsai and our colleagues for their contributions to this project. Katie, Indy, Grampa, and Desmond contributed invaluably to the inspiration and development of the pet apps. Funding was provided by Snap Inc. The first author is also supported by a NASA grant (80NSSC19K1133).

\end{acks}



\begin{thebibliography}{81}


\ifx \showCODEN    \undefined \def \showCODEN     #1{\unskip}     \fi
\ifx \showDOI      \undefined \def \showDOI       #1{#1}\fi
\ifx \showISBNx    \undefined \def \showISBNx     #1{\unskip}     \fi
\ifx \showISBNxiii \undefined \def \showISBNxiii  #1{\unskip}     \fi
\ifx \showISSN     \undefined \def \showISSN      #1{\unskip}     \fi
\ifx \showLCCN     \undefined \def \showLCCN      #1{\unskip}     \fi
\ifx \shownote     \undefined \def \shownote      #1{#1}          \fi
\ifx \showarticletitle \undefined \def \showarticletitle #1{#1}   \fi
\ifx \showURL      \undefined \def \showURL       {\relax}        \fi
\providecommand\bibfield[2]{#2}
\providecommand\bibinfo[2]{#2}
\providecommand\natexlab[1]{#1}
\providecommand\showeprint[2][]{arXiv:#2}

\bibitem[\protect\citeauthoryear{Abowd, Dey, Brown, Davies, Smith, and
  Steggles}{Abowd et~al\mbox{.}}{1999}]%
        {abowd1999towards}
\bibfield{author}{\bibinfo{person}{Gregory~D Abowd}, \bibinfo{person}{Anind~K
  Dey}, \bibinfo{person}{Peter~J Brown}, \bibinfo{person}{Nigel Davies},
  \bibinfo{person}{Mark Smith}, {and} \bibinfo{person}{Pete Steggles}.}
  \bibinfo{year}{1999}\natexlab{}.
\newblock \showarticletitle{Towards a Better Understanding of Context and
  Context-Awareness}. In \bibinfo{booktitle}{\emph{International Symposium on
  Handheld and Ubiquitous Computing}}. Springer, \bibinfo{pages}{304--307}.
\newblock


\bibitem[\protect\citeauthoryear{Alharthi, Spiel, Hamilton, Bonsignore, and
  Toups}{Alharthi et~al\mbox{.}}{2018}]%
        {alharthi2018collaborative}
\bibfield{author}{\bibinfo{person}{Sultan~A Alharthi}, \bibinfo{person}{Katta
  Spiel}, \bibinfo{person}{William~A Hamilton}, \bibinfo{person}{Elizabeth
  Bonsignore}, {and} \bibinfo{person}{Zachary~O Toups}.}
  \bibinfo{year}{2018}\natexlab{}.
\newblock \showarticletitle{Collaborative Mixed Reality Games}. In
  \bibinfo{booktitle}{\emph{Companion of the 2018 ACM Conference on Computer
  Supported Cooperative Work and Social Computing}}. \bibinfo{pages}{447--454}.
\newblock


\bibitem[\protect\citeauthoryear{Angus}{Angus}{2017}]%
        {21Tweets5:online}
\bibfield{author}{\bibinfo{person}{Kat Angus}.}
  \bibinfo{year}{2017}\natexlab{}.
\newblock \showarticletitle{21 Tweets That Prove That Cats Are Adorable
  Assholes You Can't Help But Love}.
\newblock
  \bibinfo{howpublished}{\url{https://www.buzzfeed.com/katangus/cat-owner-tweets}}.
\newblock \bibinfo{journal}{\emph{BuzzFeed}} (\bibinfo{date}{September}
  \bibinfo{year}{2017}).
\newblock
\newblock
\shownote{(Accessed on 02/04/2023).}


\bibitem[\protect\citeauthoryear{Balbus}{Balbus}{2019}]%
        {balbus_2019}
\bibfield{author}{\bibinfo{person}{Skyler Balbus}.}
  \bibinfo{year}{2019}\natexlab{}.
\newblock \bibinfo{title}{Screen time, sacred time}.
\newblock
\newblock
\urldef\tempurl%
\url{https://reallifemag.com/screen-time-sacred-time/}
\showURL{%
\tempurl}


\bibitem[\protect\citeauthoryear{Bandai}{Bandai}{1996}]%
        {tamagotchi}
\bibfield{author}{\bibinfo{person}{Bandai}.} \bibinfo{year}{1996}\natexlab{}.
\newblock \bibinfo{title}{Tamagotchi}.
\newblock
\newblock
\urldef\tempurl%
\url{https://tamagotchi.com/}
\showURL{%
\tempurl}
\newblock
\shownote{(Accessed on 01/10/2022).}


\bibitem[\protect\citeauthoryear{Bartle}{Bartle}{2004}]%
        {bartle2004designing}
\bibfield{author}{\bibinfo{person}{Richard~A Bartle}.}
  \bibinfo{year}{2004}\natexlab{}.
\newblock \bibinfo{booktitle}{\emph{Designing Virtual Worlds}}.
\newblock \bibinfo{publisher}{New Riders}.
\newblock


\bibitem[\protect\citeauthoryear{Bennett, Brady, and Branham}{Bennett
  et~al\mbox{.}}{2018}]%
        {Bennett2018}
\bibfield{author}{\bibinfo{person}{Cynthia~L. Bennett}, \bibinfo{person}{Erin
  Brady}, {and} \bibinfo{person}{Stacy~M. Branham}.}
  \bibinfo{year}{2018}\natexlab{}.
\newblock \showarticletitle{Interdependence as a {Frame} for {Assistive}
  {Technology} {Research} and {Design}}. In
  \bibinfo{booktitle}{\emph{Proceedings of the 20th {International} {ACM}
  {SIGACCESS} {Conference} on {Computers} and {Accessibility}}}.
  \bibinfo{publisher}{ACM}, \bibinfo{address}{Galway Ireland},
  \bibinfo{pages}{161--173}.
\newblock
\showISBNx{978-1-4503-5650-3}
\urldef\tempurl%
\url{https://doi.org/10.1145/3234695.3236348}
\showDOI{\tempurl}


\bibitem[\protect\citeauthoryear{Beyer and Holtzblatt}{Beyer and
  Holtzblatt}{1999}]%
        {beyer1999contextual}
\bibfield{author}{\bibinfo{person}{Hugh Beyer} {and} \bibinfo{person}{Karen
  Holtzblatt}.} \bibinfo{year}{1999}\natexlab{}.
\newblock \showarticletitle{Contextual Design}.
\newblock \bibinfo{journal}{\emph{Interactions}} \bibinfo{volume}{6},
  \bibinfo{number}{1} (\bibinfo{year}{1999}), \bibinfo{pages}{32--42}.
\newblock


\bibitem[\protect\citeauthoryear{{BoardGameGeek, LLC}}{{BoardGameGeek,
  LLC}}{2019}]%
        {boardgamegeek}
\bibfield{author}{\bibinfo{person}{{BoardGameGeek, LLC}}.}
  \bibinfo{year}{2019}\natexlab{}.
\newblock \bibinfo{title}{Escape room in a box: Flashback}.
\newblock
\newblock
\urldef\tempurl%
\url{https://boardgamegeek.com/boardgame/279911/escape-room-box-flashback}
\showURL{%
\tempurl}
\newblock
\shownote{(Accessed on 02/04/2023).}


\bibitem[\protect\citeauthoryear{Broom, Lee, Lam, and Flint}{Broom
  et~al\mbox{.}}{2019}]%
        {broom2019go}
\bibfield{author}{\bibinfo{person}{David~Robert Broom}, \bibinfo{person}{Ka~Yiu
  Lee}, \bibinfo{person}{Michael Huen~Sum Lam}, {and}
  \bibinfo{person}{Stuart~William Flint}.} \bibinfo{year}{2019}\natexlab{}.
\newblock \showarticletitle{Go ta catch 'em al or not enough time: Users
  motivations for playing Pok{\'e}mon Go™ and non-users’ reasons for not
  installing}.
\newblock \bibinfo{journal}{\emph{Health Psychology Research}}
  \bibinfo{volume}{7}, \bibinfo{number}{1} (\bibinfo{year}{2019}).
\newblock


\bibitem[\protect\citeauthoryear{Brown and Ryan}{Brown and Ryan}{2003}]%
        {brown2003benefits}
\bibfield{author}{\bibinfo{person}{Kirk~Warren Brown} {and}
  \bibinfo{person}{Richard~M Ryan}.} \bibinfo{year}{2003}\natexlab{}.
\newblock \showarticletitle{The benefits of being present: mindfulness and its
  role in psychological well-being.}
\newblock \bibinfo{journal}{\emph{Journal of personality and social
  psychology}} \bibinfo{volume}{84}, \bibinfo{number}{4}
  (\bibinfo{year}{2003}), \bibinfo{pages}{822}.
\newblock


\bibitem[\protect\citeauthoryear{Burrell}{Burrell}{2010}]%
        {burrell2010evaluating}
\bibfield{author}{\bibinfo{person}{Jenna Burrell}.}
  \bibinfo{year}{2010}\natexlab{}.
\newblock \showarticletitle{Evaluating Shared Access: social equality and the
  circulation of mobile phones in rural Uganda}.
\newblock \bibinfo{journal}{\emph{Journal of Computer-Mediated Communication}}
  \bibinfo{volume}{15}, \bibinfo{number}{2} (\bibinfo{year}{2010}),
  \bibinfo{pages}{230--250}.
\newblock


\bibitem[\protect\citeauthoryear{Chan}{Chan}{2015}]%
        {chan2015multimodal}
\bibfield{author}{\bibinfo{person}{Michael Chan}.}
  \bibinfo{year}{2015}\natexlab{}.
\newblock \showarticletitle{Multimodal connectedness and quality of life:
  Examining the influences of technology adoption and interpersonal
  communication on well-being across the life span}.
\newblock \bibinfo{journal}{\emph{Journal of Computer-Mediated Communication}}
  \bibinfo{volume}{20}, \bibinfo{number}{1} (\bibinfo{year}{2015}),
  \bibinfo{pages}{3--18}.
\newblock


\bibitem[\protect\citeauthoryear{Chen, Liu, Li, Yu, Gao, Caon, Yue, and
  Liang}{Chen et~al\mbox{.}}{2021}]%
        {chen2021effect}
\bibfield{author}{\bibinfo{person}{Lei Chen}, \bibinfo{person}{Yilin Liu},
  \bibinfo{person}{Yue Li}, \bibinfo{person}{Lingyun Yu}, \bibinfo{person}{BoYu
  Gao}, \bibinfo{person}{Maurizio Caon}, \bibinfo{person}{Yong Yue}, {and}
  \bibinfo{person}{Hai-Ning Liang}.} \bibinfo{year}{2021}\natexlab{}.
\newblock \showarticletitle{Effect of visual cues on pointing tasks in
  co-located augmented reality collaboration}. In
  \bibinfo{booktitle}{\emph{Symposium on Spatial User Interaction}}.
  \bibinfo{pages}{1--12}.
\newblock


\bibitem[\protect\citeauthoryear{Cheng, Yan, Yi, Shi, and Lindlbauer}{Cheng
  et~al\mbox{.}}{2021}]%
        {semanticadapt}
\bibfield{author}{\bibinfo{person}{Yifei Cheng}, \bibinfo{person}{Yukang Yan},
  \bibinfo{person}{Xin Yi}, \bibinfo{person}{Yuanchun Shi}, {and}
  \bibinfo{person}{David Lindlbauer}.} \bibinfo{year}{2021}\natexlab{}.
\newblock \showarticletitle{SemanticAdapt: Optimization-based Adaptation of
  Mixed Reality Layouts Leveraging Virtual-Physical Semantic Connections}. In
  \bibinfo{booktitle}{\emph{Proceedings of the 34th Annual Symposium on User
  Interface Software and Technology}}. \bibinfo{pages}{282--297}.
\newblock


\bibitem[\protect\citeauthoryear{Chittaro and Sioni}{Chittaro and
  Sioni}{2014}]%
        {CHITTARO201456}
\bibfield{author}{\bibinfo{person}{Luca Chittaro} {and}
  \bibinfo{person}{Riccardo Sioni}.} \bibinfo{year}{2014}\natexlab{}.
\newblock \showarticletitle{Evaluating mobile apps for breathing training: The
  effectiveness of visualization}.
\newblock \bibinfo{journal}{\emph{Computers in Human Behavior}}
  \bibinfo{volume}{40} (\bibinfo{year}{2014}), \bibinfo{pages}{56--63}.
\newblock
\showISSN{0747-5632}
\urldef\tempurl%
\url{https://doi.org/10.1016/j.chb.2014.07.049}
\showDOI{\tempurl}


\bibitem[\protect\citeauthoryear{Chotpitayasunondh and
  Douglas}{Chotpitayasunondh and Douglas}{2016}]%
        {chotpitayasunondh2016phubbing}
\bibfield{author}{\bibinfo{person}{Varoth Chotpitayasunondh} {and}
  \bibinfo{person}{Karen~M Douglas}.} \bibinfo{year}{2016}\natexlab{}.
\newblock \showarticletitle{How “phubbing” becomes the norm: The
  antecedents and consequences of snubbing via smartphone}.
\newblock \bibinfo{journal}{\emph{Computers in Human Behavior}}
  \bibinfo{volume}{63} (\bibinfo{year}{2016}), \bibinfo{pages}{9--18}.
\newblock


\bibitem[\protect\citeauthoryear{Chotpitayasunondh and
  Douglas}{Chotpitayasunondh and Douglas}{2018}]%
        {chotpitayasunondh2018effects}
\bibfield{author}{\bibinfo{person}{Varoth Chotpitayasunondh} {and}
  \bibinfo{person}{Karen~M Douglas}.} \bibinfo{year}{2018}\natexlab{}.
\newblock \showarticletitle{The effects of “phubbing” on social
  interaction}.
\newblock \bibinfo{journal}{\emph{Journal of Applied Social Psychology}}
  \bibinfo{volume}{48}, \bibinfo{number}{6} (\bibinfo{year}{2018}),
  \bibinfo{pages}{304--316}.
\newblock


\bibitem[\protect\citeauthoryear{Dagan, C\'{a}rdenas~Gasca, Robinson, Noriega,
  Tham, Vaish, and Monroy-Hern\'{a}ndez}{Dagan et~al\mbox{.}}{2022}]%
        {irl_2022}
\bibfield{author}{\bibinfo{person}{Ella Dagan}, \bibinfo{person}{Ana~Mar\'{\i}a
  C\'{a}rdenas~Gasca}, \bibinfo{person}{Ava Robinson}, \bibinfo{person}{Anwar
  Noriega}, \bibinfo{person}{Yu~Jiang Tham}, \bibinfo{person}{Rajan Vaish},
  {and} \bibinfo{person}{Andr\'{e}s Monroy-Hern\'{a}ndez}.}
  \bibinfo{year}{2022}\natexlab{}.
\newblock \showarticletitle{Project IRL: Playful Co-Located Interactions with
  Mobile Augmented Reality}.
\newblock \bibinfo{journal}{\emph{Proc. ACM Hum.-Comput. Interact.}}
  \bibinfo{volume}{6}, \bibinfo{number}{CSCW1}, Article \bibinfo{articleno}{62}
  (\bibinfo{date}{apr} \bibinfo{year}{2022}), \bibinfo{numpages}{27}~pages.
\newblock
\urldef\tempurl%
\url{https://doi.org/10.1145/3512909}
\showDOI{\tempurl}


\bibitem[\protect\citeauthoryear{Daud\'{e}n~Roquet and Sas}{Daud\'{e}n~Roquet
  and Sas}{2018}]%
        {dauden2018review}
\bibfield{author}{\bibinfo{person}{Claudia Daud\'{e}n~Roquet} {and}
  \bibinfo{person}{Corina Sas}.} \bibinfo{year}{2018}\natexlab{}.
\newblock \showarticletitle{Evaluating Mindfulness Meditation Apps}. In
  \bibinfo{booktitle}{\emph{Extended Abstracts of the 2018 CHI Conference on
  Human Factors in Computing Systems}} (Montreal QC, Canada)
  \emph{(\bibinfo{series}{CHI EA '18})}. \bibinfo{publisher}{Association for
  Computing Machinery}, \bibinfo{address}{New York, NY, USA},
  \bibinfo{pages}{1–6}.
\newblock
\showISBNx{9781450356213}
\urldef\tempurl%
\url{https://doi.org/10.1145/3170427.3188616}
\showDOI{\tempurl}


\bibitem[\protect\citeauthoryear{Davis}{Davis}{2016}]%
        {davis_2016}
\bibfield{author}{\bibinfo{person}{Jenny~L. Davis}.}
  \bibinfo{year}{2016}\natexlab{}.
\newblock \bibinfo{title}{Face to interface}.
\newblock
\newblock
\urldef\tempurl%
\url{https://reallifemag.com/face-to-interface/}
\showURL{%
\tempurl}


\bibitem[\protect\citeauthoryear{doglab}{doglab}{[n.d.]}]%
        {doglab}
doglab \bibinfo{year}{[n.d.]}\natexlab{}.
\newblock \bibinfo{title}{Canine Cognition Center}.
\newblock
\newblock
\urldef\tempurl%
\url{https://doglab.yale.edu/}
\showURL{%
\tempurl}


\bibitem[\protect\citeauthoryear{Erg{\"u}n, G{\"o}ksu, and Sak{\i}z}{Erg{\"u}n
  et~al\mbox{.}}{2020}]%
        {ergun2020effects}
\bibfield{author}{\bibinfo{person}{Naif Erg{\"u}n}, \bibinfo{person}{{\.I}dris
  G{\"o}ksu}, {and} \bibinfo{person}{Halis Sak{\i}z}.}
  \bibinfo{year}{2020}\natexlab{}.
\newblock \showarticletitle{Effects of phubbing: Relationships with
  psychodemographic variables}.
\newblock \bibinfo{journal}{\emph{Psychological Reports}}
  \bibinfo{volume}{123}, \bibinfo{number}{5} (\bibinfo{year}{2020}),
  \bibinfo{pages}{1578--1613}.
\newblock


\bibitem[\protect\citeauthoryear{Fan, Li, Zhong, Tian, Shi, and Wang}{Fan
  et~al\mbox{.}}{2011}]%
        {fan2011surprise}
\bibfield{author}{\bibinfo{person}{Mingming Fan}, \bibinfo{person}{Xin Li},
  \bibinfo{person}{Yu Zhong}, \bibinfo{person}{Li Tian},
  \bibinfo{person}{Yuanchun Shi}, {and} \bibinfo{person}{Hao Wang}.}
  \bibinfo{year}{2011}\natexlab{}.
\newblock \showarticletitle{Surprise Grabber: a co-located tangible social game
  using phone hand gesture}. In \bibinfo{booktitle}{\emph{Proceedings of the
  ACM 2011 Conference on Computer Supported Cooperative Work}}.
  \bibinfo{pages}{625--628}.
\newblock


\bibitem[\protect\citeauthoryear{Fischer, Porcheron, Lucero, Quigley, Scott,
  Ciolfi, Rooksby, and Memarovic}{Fischer et~al\mbox{.}}{2016}]%
        {fischer2016collocated}
\bibfield{author}{\bibinfo{person}{Joel Fischer}, \bibinfo{person}{Martin
  Porcheron}, \bibinfo{person}{Andr{\'e}s Lucero}, \bibinfo{person}{Aaron
  Quigley}, \bibinfo{person}{Stacey Scott}, \bibinfo{person}{Luigina Ciolfi},
  \bibinfo{person}{John Rooksby}, {and} \bibinfo{person}{Nemanja Memarovic}.}
  \bibinfo{year}{2016}\natexlab{}.
\newblock \showarticletitle{Collocated interaction: new challenges in `same
  time, same place' research}. In \bibinfo{booktitle}{\emph{Proceedings of the
  19th ACM Conference on Computer Supported Cooperative Work and Social
  Computing Companion}}. \bibinfo{pages}{465--472}.
\newblock


\bibitem[\protect\citeauthoryear{Grevet and Gilbert}{Grevet and
  Gilbert}{2015}]%
        {grevet2015piggyback}
\bibfield{author}{\bibinfo{person}{Catherine Grevet} {and}
  \bibinfo{person}{Eric Gilbert}.} \bibinfo{year}{2015}\natexlab{}.
\newblock \showarticletitle{Piggyback prototyping: Using existing, large-scale
  social computing systems to prototype new ones}. In
  \bibinfo{booktitle}{\emph{Proceedings of the 33rd Annual ACM Conference on
  Human Factors in Computing Systems}}. \bibinfo{pages}{4047--4056}.
\newblock


\bibitem[\protect\citeauthoryear{Hassenzahl, Heidecker, Eckoldt, Diefenbach,
  and Hillmann}{Hassenzahl et~al\mbox{.}}{2012}]%
        {hassenzahl2012allyouneed}
\bibfield{author}{\bibinfo{person}{Marc Hassenzahl}, \bibinfo{person}{Stephanie
  Heidecker}, \bibinfo{person}{Kai Eckoldt}, \bibinfo{person}{Sarah
  Diefenbach}, {and} \bibinfo{person}{Uwe Hillmann}.}
  \bibinfo{year}{2012}\natexlab{}.
\newblock \showarticletitle{All You Need is Love: Current Strategies of
  Mediating Intimate Relationships through Technology}.
\newblock \bibinfo{journal}{\emph{ACM Trans. Comput.-Hum. Interact.}}
  \bibinfo{volume}{19}, \bibinfo{number}{4}, Article \bibinfo{articleno}{30}
  (\bibinfo{date}{Dec} \bibinfo{year}{2012}), \bibinfo{numpages}{19}~pages.
\newblock
\showISSN{1073-0516}
\urldef\tempurl%
\url{https://doi.org/10.1145/2395131.2395137}
\showDOI{\tempurl}


\bibitem[\protect\citeauthoryear{Hauser, Wakkary, and Neustaedter}{Hauser
  et~al\mbox{.}}{2014}]%
        {hauser2014improving}
\bibfield{author}{\bibinfo{person}{Sabrina Hauser}, \bibinfo{person}{Ron
  Wakkary}, {and} \bibinfo{person}{Carman Neustaedter}.}
  \bibinfo{year}{2014}\natexlab{}.
\newblock \showarticletitle{Improving guide dog team play with accessible dog
  toys}.
\newblock In \bibinfo{booktitle}{\emph{CHI'14 Extended Abstracts on Human
  Factors in Computing Systems}}. \bibinfo{pages}{1537--1542}.
\newblock


\bibitem[\protect\citeauthoryear{Henderson and Feiner}{Henderson and
  Feiner}{2010}]%
        {opportunistic}
\bibfield{author}{\bibinfo{person}{Steven Henderson} {and}
  \bibinfo{person}{Steven Feiner}.} \bibinfo{year}{2010}\natexlab{}.
\newblock \showarticletitle{Opportunistic Tangible User Interfaces for
  Augmented Reality}.
\newblock \bibinfo{journal}{\emph{IEEE Transactions on Visualization and
  Computer Graphics}} \bibinfo{volume}{16}, \bibinfo{number}{1}
  (\bibinfo{year}{2010}).
\newblock


\bibitem[\protect\citeauthoryear{Hirskyj-Douglas and Lucero}{Hirskyj-Douglas
  and Lucero}{2021}]%
        {hirskyj2021forming}
\bibfield{author}{\bibinfo{person}{Ilyena Hirskyj-Douglas} {and}
  \bibinfo{person}{Andr{\'e}s Lucero}.} \bibinfo{year}{2021}\natexlab{}.
\newblock \showarticletitle{Forming the Dog Internet: Prototyping a
  Dog-to-Human Video Call Device}.
\newblock \bibinfo{journal}{\emph{Proc. ACM Hum.-Comput. Interact.}}
  \bibinfo{volume}{5}, \bibinfo{number}{ISS} (\bibinfo{year}{2021}).
\newblock


\bibitem[\protect\citeauthoryear{Hirskyj-Douglas and Read}{Hirskyj-Douglas and
  Read}{2016}]%
        {hirskyjread2016ethics}
\bibfield{author}{\bibinfo{person}{Ilyena Hirskyj-Douglas} {and}
  \bibinfo{person}{Janet~C. Read}.} \bibinfo{year}{2016}\natexlab{}.
\newblock \showarticletitle{The ethics of how to work with dogs in animal
  computer interaction}. In \bibinfo{booktitle}{\emph{Proceedings of Measuring
  Behavior 2016, (Dublin, Ireland, 25-27 May 2016)}}.
\newblock


\bibitem[\protect\citeauthoryear{HoloLens}{HoloLens}{[n.d.]}]%
        {hololens}
HoloLens \bibinfo{year}{[n.d.]}\natexlab{}.
\newblock \bibinfo{title}{Microsoft HoloLens: Mixed Reality Technology for
  Business}.
\newblock
\newblock
\urldef\tempurl%
\url{https://www.microsoft.com/en-us/hololens}
\showURL{%
\tempurl}


\bibitem[\protect\citeauthoryear{Howard, Zhu, Chen, Kalenichenko, Wang, Weyand,
  Andreetto, and Adam}{Howard et~al\mbox{.}}{2017}]%
        {howard2017mobilenets}
\bibfield{author}{\bibinfo{person}{Andrew~G Howard}, \bibinfo{person}{Menglong
  Zhu}, \bibinfo{person}{Bo Chen}, \bibinfo{person}{Dmitry Kalenichenko},
  \bibinfo{person}{Weijun Wang}, \bibinfo{person}{Tobias Weyand},
  \bibinfo{person}{Marco Andreetto}, {and} \bibinfo{person}{Hartwig Adam}.}
  \bibinfo{year}{2017}\natexlab{}.
\newblock \showarticletitle{Mobilenets: Efficient convolutional neural networks
  for mobile vision applications}.
\newblock \bibinfo{journal}{\emph{arXiv preprint arXiv:1704.04861}}
  (\bibinfo{year}{2017}).
\newblock


\bibitem[\protect\citeauthoryear{Jarusriboonchai, Malapaschas, and
  Olsson}{Jarusriboonchai et~al\mbox{.}}{2016}]%
        {jarusriboonchai2016design}
\bibfield{author}{\bibinfo{person}{Pradthana Jarusriboonchai},
  \bibinfo{person}{Aris Malapaschas}, {and} \bibinfo{person}{Thomas Olsson}.}
  \bibinfo{year}{2016}\natexlab{}.
\newblock \showarticletitle{Design and evaluation of a multi-player mobile game
  for icebreaking activity}. In \bibinfo{booktitle}{\emph{Proceedings of the
  2016 CHI Conference on Human Factors in Computing Systems}}.
  \bibinfo{pages}{4366--4377}.
\newblock


\bibitem[\protect\citeauthoryear{Joffe}{Joffe}{2012}]%
        {joffe2012thematic}
\bibfield{author}{\bibinfo{person}{Helene Joffe}.}
  \bibinfo{year}{2012}\natexlab{}.
\newblock \showarticletitle{Thematic Analysis}.
\newblock \bibinfo{journal}{\emph{Qualitative Research Methods in Mental Health
  and Psychotherapy}}  \bibinfo{volume}{1} (\bibinfo{year}{2012}).
\newblock


\bibitem[\protect\citeauthoryear{Kalarchian, Hammer, and
  Kapuścińska}{Kalarchian et~al\mbox{.}}{2021}]%
        {Frolic:online}
\bibfield{author}{\bibinfo{person}{Melissa~A Kalarchian},
  \bibinfo{person}{Jessica Hammer}, {and} \bibinfo{person}{Adela
  Kapuścińska}.} \bibinfo{year}{2021}\natexlab{}.
\newblock \showarticletitle{Fostering Innovation in Prevention and Treatment of
  Obesity in Youth: Digitally Mediated Physical Play as an Exemplar}.
\newblock \bibinfo{journal}{\emph{Obesity}} \bibinfo{volume}{29},
  \bibinfo{number}{3} (\bibinfo{year}{2021}), \bibinfo{pages}{475--477}.
\newblock


\bibitem[\protect\citeauthoryear{Kardefelt-Winther, Rees, and
  Livingstone}{Kardefelt-Winther et~al\mbox{.}}{2020}]%
        {kardefelt2020contextualising}
\bibfield{author}{\bibinfo{person}{Daniel Kardefelt-Winther},
  \bibinfo{person}{Gwyther Rees}, {and} \bibinfo{person}{Sonia Livingstone}.}
  \bibinfo{year}{2020}\natexlab{}.
\newblock \showarticletitle{Contextualising the link between adolescents’ use
  of digital technology and their mental health: a multi-country study of time
  spent online and life satisfaction}.
\newblock \bibinfo{journal}{\emph{Journal of Child Psychology and Psychiatry}}
  \bibinfo{volume}{61}, \bibinfo{number}{8} (\bibinfo{year}{2020}),
  \bibinfo{pages}{875--889}.
\newblock


\bibitem[\protect\citeauthoryear{Kilteni, Groten, and Slater}{Kilteni
  et~al\mbox{.}}{2012}]%
        {embodiment:online}
\bibfield{author}{\bibinfo{person}{Konstantina Kilteni},
  \bibinfo{person}{Raphaela Groten}, {and} \bibinfo{person}{Mel Slater}.}
  \bibinfo{year}{2012}\natexlab{}.
\newblock \showarticletitle{The Sense of Embodiment in Virtual Reality}.
\newblock \bibinfo{journal}{\emph{Presence: Teleoperators and Virtual
  Environments}} \bibinfo{volume}{21}, \bibinfo{number}{4}
  (\bibinfo{year}{2012}), \bibinfo{pages}{373--387}.
\newblock


\bibitem[\protect\citeauthoryear{Liu, Dabbish, and Kaufman}{Liu
  et~al\mbox{.}}{2017}]%
        {liu2017supporting}
\bibfield{author}{\bibinfo{person}{Fannie Liu}, \bibinfo{person}{Laura
  Dabbish}, {and} \bibinfo{person}{Geoff Kaufman}.}
  \bibinfo{year}{2017}\natexlab{}.
\newblock \showarticletitle{Supporting social interactions with an expressive
  heart rate sharing application}.
\newblock \bibinfo{journal}{\emph{Proceedings of the ACM on Interactive,
  Mobile, Wearable and Ubiquitous Technologies}} \bibinfo{volume}{1},
  \bibinfo{number}{3} (\bibinfo{year}{2017}), \bibinfo{pages}{1--26}.
\newblock


\bibitem[\protect\citeauthoryear{Liu, Smith, Vaish, and
  Monroy-Hern{\'a}ndez}{Liu et~al\mbox{.}}{2022}]%
        {liu2022}
\bibfield{author}{\bibinfo{person}{Szu-Yu Liu}, \bibinfo{person}{Brian~A
  Smith}, \bibinfo{person}{Rajan Vaish}, {and} \bibinfo{person}{Andr{\'e}s
  Monroy-Hern{\'a}ndez}.} \bibinfo{year}{2022}\natexlab{}.
\newblock \bibinfo{journal}{\emph{Proceedings of the ACM on Human-Computer
  Interaction}} \bibinfo{volume}{6}, \bibinfo{number}{CSCW1}
  (\bibinfo{year}{2022}), \bibinfo{pages}{1--26}.
\newblock


\bibitem[\protect\citeauthoryear{Lucero, Holopainen, and Jokela}{Lucero
  et~al\mbox{.}}{2011}]%
        {lucero2011pass}
\bibfield{author}{\bibinfo{person}{Andr{\'e}s Lucero}, \bibinfo{person}{Jussi
  Holopainen}, {and} \bibinfo{person}{Tero Jokela}.}
  \bibinfo{year}{2011}\natexlab{}.
\newblock \showarticletitle{Pass-them-around: collaborative use of mobile
  phones for photo sharing}. In \bibinfo{booktitle}{\emph{Proceedings of the
  SIGCHI COnference on Human Factors in Computing Systems}}.
  \bibinfo{pages}{1787--1796}.
\newblock


\bibitem[\protect\citeauthoryear{Lucero, Ker{\"a}nen, and Jokela}{Lucero
  et~al\mbox{.}}{2010a}]%
        {lucero2010social}
\bibfield{author}{\bibinfo{person}{Andr{\'e}s Lucero}, \bibinfo{person}{Jaakko
  Ker{\"a}nen}, {and} \bibinfo{person}{Tero Jokela}.}
  \bibinfo{year}{2010}\natexlab{a}.
\newblock \showarticletitle{Social and spatial interactions: shared co-located
  mobile phone use}.
\newblock In \bibinfo{booktitle}{\emph{CHI'10 extended abstracts on human
  factors in computing systems}}. \bibinfo{pages}{3223--3228}.
\newblock


\bibitem[\protect\citeauthoryear{Lucero, Ker{\"a}nen, and Korhonen}{Lucero
  et~al\mbox{.}}{2010b}]%
        {lucero2010collaborative}
\bibfield{author}{\bibinfo{person}{Andr{\'e}s Lucero}, \bibinfo{person}{Jaakko
  Ker{\"a}nen}, {and} \bibinfo{person}{Hannu Korhonen}.}
  \bibinfo{year}{2010}\natexlab{b}.
\newblock \showarticletitle{Collaborative use of mobile phones for
  brainstorming}. In \bibinfo{booktitle}{\emph{Proceedings of the 12th
  international conference on Human computer interaction with mobile devices
  and services}}. \bibinfo{pages}{337--340}.
\newblock


\bibitem[\protect\citeauthoryear{Magic Leap}{Magic Leap}{[n.d.]}]%
        {magicleap}
Magic Leap \bibinfo{year}{[n.d.]}\natexlab{}.
\newblock \bibinfo{title}{Enterprise augmented reality (AR) platform designed
  for business: Magic Leap}.
\newblock
\newblock
\urldef\tempurl%
\url{https://www.magicleap.com/en-us/}
\showURL{%
\tempurl}


\bibitem[\protect\citeauthoryear{Memarovic, Langheinrich, Kostakos,
  Fitzpatrick, and Huang}{Memarovic et~al\mbox{.}}{2012}]%
        {memarovic2012workshop}
\bibfield{author}{\bibinfo{person}{Nemanja Memarovic}, \bibinfo{person}{Marc
  Langheinrich}, \bibinfo{person}{Vassilis Kostakos},
  \bibinfo{person}{Geraldine Fitzpatrick}, {and} \bibinfo{person}{Elaine~M
  Huang}.} \bibinfo{year}{2012}\natexlab{}.
\newblock \showarticletitle{Workshop on Computer Mediated Social Offline
  Interactions (SOFTec 2012)}. In \bibinfo{booktitle}{\emph{Proceedings of the
  2012 ACM Conference on Ubiquitous Computing}}. \bibinfo{pages}{790--791}.
\newblock


\bibitem[\protect\citeauthoryear{Nair, Ma, Huddleston, Lin, Hayes, Donnelly,
  Penuela, He, and Smith}{Nair et~al\mbox{.}}{2021}]%
        {AcousticMinimap:online}
\bibfield{author}{\bibinfo{person}{Vishnu Nair}, \bibinfo{person}{Shao-en Ma},
  \bibinfo{person}{Hannah Huddleston}, \bibinfo{person}{Karen Lin},
  \bibinfo{person}{Mason Hayes}, \bibinfo{person}{Matthew Donnelly},
  \bibinfo{person}{Ricardo E.~Gonzalez Penuela}, \bibinfo{person}{Yicheng He},
  {and} \bibinfo{person}{Brian~A. Smith}.} \bibinfo{year}{2021}\natexlab{}.
\newblock \showarticletitle{Towards a Generalized Acoustic Minimap for Visually
  Impaired Gamers}. In \bibinfo{booktitle}{\emph{UIST '21: The Adjunct
  Publication of the 34th Annual ACM Symposium on User Interface Software and
  Technology}}. \bibinfo{pages}{89--91}.
\newblock


\bibitem[\protect\citeauthoryear{Nicholas, Smith, and Vaish}{Nicholas
  et~al\mbox{.}}{2021}]%
        {Friendscope:online}
\bibfield{author}{\bibinfo{person}{Molly~Jane Nicholas},
  \bibinfo{person}{Brian~A. Smith}, {and} \bibinfo{person}{Rajan Vaish}.}
  \bibinfo{year}{2021}\natexlab{}.
\newblock \showarticletitle{Friendscope: Exploring In-the-Moment Experience
  Sharing on Camera Glasses via a Shared Camera}. In
  \bibinfo{booktitle}{\emph{Proceedings of ACM Human-Computer Interaction, CSCW
  Issue 2022}}. \bibinfo{pages}{1--25}.
\newblock


\bibitem[\protect\citeauthoryear{Object Tracking}{Object Tracking}{[n.d.]}]%
        {objecttrackingdocs}
Object Tracking \bibinfo{year}{[n.d.]}\natexlab{}.
\newblock \bibinfo{title}{Object Tracking}.
\newblock
\newblock
\urldef\tempurl%
\url{https://docs.snap.com/lens-studio/references/guides/lens-features/tracking/world/object-tracking/}
\showURL{%
\tempurl}


\bibitem[\protect\citeauthoryear{Odom, Zimmerman, and Forlizzi}{Odom
  et~al\mbox{.}}{2011}]%
        {odom2011teenagers}
\bibfield{author}{\bibinfo{person}{William Odom}, \bibinfo{person}{John
  Zimmerman}, {and} \bibinfo{person}{Jodi Forlizzi}.}
  \bibinfo{year}{2011}\natexlab{}.
\newblock \showarticletitle{Teenagers and their virtual possessions: design
  opportunities and issues}. In \bibinfo{booktitle}{\emph{Proceedings of the
  SIGCHI Conference on Human Factors in Computing Systems}}.
  \bibinfo{pages}{1491--1500}.
\newblock


\bibitem[\protect\citeauthoryear{Olsson, Jarusriboonchai, Wo{\'z}niak,
  Paasovaara, V{\"a}{\"a}n{\"a}nen, and Lucero}{Olsson et~al\mbox{.}}{2020}]%
        {olsson2020technologies}
\bibfield{author}{\bibinfo{person}{Thomas Olsson}, \bibinfo{person}{Pradthana
  Jarusriboonchai}, \bibinfo{person}{Pawe{\l} Wo{\'z}niak},
  \bibinfo{person}{Susanna Paasovaara}, \bibinfo{person}{Kaisa
  V{\"a}{\"a}n{\"a}nen}, {and} \bibinfo{person}{Andr{\'e}s Lucero}.}
  \bibinfo{year}{2020}\natexlab{}.
\newblock \showarticletitle{Technologies for enhancing collocated social
  interaction: review of design solutions and approaches}.
\newblock \bibinfo{journal}{\emph{Computer Supported Cooperative Work (CSCW)}}
  \bibinfo{volume}{29}, \bibinfo{number}{1} (\bibinfo{year}{2020}),
  \bibinfo{pages}{29--83}.
\newblock


\bibitem[\protect\citeauthoryear{Orth, Thurgood, and Hoven}{Orth
  et~al\mbox{.}}{2019}]%
        {orth2019designing}
\bibfield{author}{\bibinfo{person}{Daniel Orth}, \bibinfo{person}{Clementine
  Thurgood}, {and} \bibinfo{person}{Elise Van~Den Hoven}.}
  \bibinfo{year}{2019}\natexlab{}.
\newblock \showarticletitle{Designing meaningful products in the digital age:
  How users value their technological possessions}.
\newblock \bibinfo{journal}{\emph{ACM Transactions on Computer-Human
  Interaction (TOCHI)}} \bibinfo{volume}{26}, \bibinfo{number}{5}
  (\bibinfo{year}{2019}), \bibinfo{pages}{1--28}.
\newblock


\bibitem[\protect\citeauthoryear{Patel}{Patel}{2021}]%
        {patel_2021}
\bibfield{author}{\bibinfo{person}{Shwetak Patel}.}
  \bibinfo{year}{2021}\natexlab{}.
\newblock \bibinfo{title}{Take a pulse on health and wellness with your phone}.
\newblock
\newblock
\urldef\tempurl%
\url{https://blog.google/technology/health/take-pulse-health-and-wellness-your-phone/}
\showURL{%
\tempurl}


\bibitem[\protect\citeauthoryear{Patibanda, Mueller, Leskovsek, and
  Duckworth}{Patibanda et~al\mbox{.}}{2017}]%
        {patibanda2017lifetree}
\bibfield{author}{\bibinfo{person}{Rakesh Patibanda},
  \bibinfo{person}{Florian~`Floyd' Mueller}, \bibinfo{person}{Matevz
  Leskovsek}, {and} \bibinfo{person}{Jonathan Duckworth}.}
  \bibinfo{year}{2017}\natexlab{}.
\newblock \showarticletitle{Life Tree: Understanding the Design of Breathing
  Exercise Games}. In \bibinfo{booktitle}{\emph{Proceedings of the Annual
  Symposium on Computer-Human Interaction in Play}} (Amsterdam, The
  Netherlands) \emph{(\bibinfo{series}{CHI PLAY '17})}.
  \bibinfo{publisher}{Association for Computing Machinery},
  \bibinfo{address}{New York, NY, USA}, \bibinfo{pages}{19–31}.
\newblock
\showISBNx{9781450348980}
\urldef\tempurl%
\url{https://doi.org/10.1145/3116595.3116621}
\showDOI{\tempurl}


\bibitem[\protect\citeauthoryear{Pei, Chen, Lee, and Zhang}{Pei
  et~al\mbox{.}}{[n.d.]}]%
        {youtubeHandInterfaces}
\bibfield{author}{\bibinfo{person}{Siyou Pei}, \bibinfo{person}{Alexander
  Chen}, \bibinfo{person}{Jaewook Lee}, {and} \bibinfo{person}{Yang Zhang}.}
  \bibinfo{year}{[n.d.]}\natexlab{}.
\newblock \bibinfo{booktitle}{\emph{Hand Interfaces: Using Hands to Imitate
  Objects in AR/VR for Expressive Interactions}}.
\newblock Youtube.
\newblock


\bibitem[\protect\citeauthoryear{Perry}{Perry}{2017}]%
        {21WeirdT32:online}
\bibfield{author}{\bibinfo{person}{Flo Perry}.}
  \bibinfo{year}{2017}\natexlab{}.
\newblock \showarticletitle{21 Weird Things Cats Do That Only Cat Owners
  Understand}.
\newblock
  \bibinfo{howpublished}{\url{https://www.buzzfeed.com/floperry/truths-about-cats-only-cat-owners-really-understand}}.
\newblock \bibinfo{journal}{\emph{BuzzFeed}} (\bibinfo{date}{September}
  \bibinfo{year}{2017}).
\newblock
\newblock
\shownote{(Accessed on 02/04/2023).}


\bibitem[\protect\citeauthoryear{Pet}{Pet}{[n.d.]}]%
        {petdocs}
Pet \bibinfo{year}{[n.d.]}\natexlab{}.
\newblock \bibinfo{title}{Pet}.
\newblock
\newblock
\urldef\tempurl%
\url{https://docs.snap.com/lens-studio/references/templates/object/pet/}
\showURL{%
\tempurl}


\bibitem[\protect\citeauthoryear{Roberts and David}{Roberts and David}{2016}]%
        {roberts2016my}
\bibfield{author}{\bibinfo{person}{James~A Roberts} {and}
  \bibinfo{person}{Meredith~E David}.} \bibinfo{year}{2016}\natexlab{}.
\newblock \showarticletitle{My life has become a major distraction from my cell
  phone: Partner phubbing and relationship satisfaction among romantic
  partners}.
\newblock \bibinfo{journal}{\emph{Computers in Human Behavior}}
  \bibinfo{volume}{54} (\bibinfo{year}{2016}), \bibinfo{pages}{134--141}.
\newblock


\bibitem[\protect\citeauthoryear{Rogers}{Rogers}{2014}]%
        {rogers2014bursting}
\bibfield{author}{\bibinfo{person}{Yvonne Rogers}.}
  \bibinfo{year}{2014}\natexlab{}.
\newblock \showarticletitle{Bursting our Digital Bubbles: Life Beyond the App}.
  In \bibinfo{booktitle}{\emph{Proceedings of the 16th International Conference
  on Multimodal Interaction}}. \bibinfo{pages}{1--1}.
\newblock


\bibitem[\protect\citeauthoryear{Roo, Gervais, Frey, and Hachet}{Roo
  et~al\mbox{.}}{2017}]%
        {roo2017inner}
\bibfield{author}{\bibinfo{person}{Joan~Sol Roo}, \bibinfo{person}{Renaud
  Gervais}, \bibinfo{person}{Jeremy Frey}, {and} \bibinfo{person}{Martin
  Hachet}.} \bibinfo{year}{2017}\natexlab{}.
\newblock \showarticletitle{Inner garden: Connecting inner states to a mixed
  reality sandbox for mindfulness}. In \bibinfo{booktitle}{\emph{Proceedings of
  the 2017 CHI Conference on Human Factors in Computing Systems}}.
  \bibinfo{pages}{1459--1470}.
\newblock


\bibitem[\protect\citeauthoryear{Rouse}{Rouse}{2020}]%
        {rouse2020you}
\bibfield{author}{\bibinfo{person}{Elizabeth~D Rouse}.}
  \bibinfo{year}{2020}\natexlab{}.
\newblock \showarticletitle{Where you end and I begin: Understanding intimate
  co-creation}.
\newblock \bibinfo{journal}{\emph{Academy of Management Review}}
  \bibinfo{volume}{45}, \bibinfo{number}{1} (\bibinfo{year}{2020}),
  \bibinfo{pages}{181--204}.
\newblock


\bibitem[\protect\citeauthoryear{Sambasivan, Checkley, Batool, Ahmed, Nemer,
  Gayt{\'a}n-Lugo, Matthews, Consolvo, and Churchill}{Sambasivan
  et~al\mbox{.}}{2018}]%
        {sambasivan2018privacy}
\bibfield{author}{\bibinfo{person}{Nithya Sambasivan}, \bibinfo{person}{Garen
  Checkley}, \bibinfo{person}{Amna Batool}, \bibinfo{person}{Nova Ahmed},
  \bibinfo{person}{David Nemer}, \bibinfo{person}{Laura~Sanely
  Gayt{\'a}n-Lugo}, \bibinfo{person}{Tara Matthews}, \bibinfo{person}{Sunny
  Consolvo}, {and} \bibinfo{person}{Elizabeth Churchill}.}
  \bibinfo{year}{2018}\natexlab{}.
\newblock \showarticletitle{``Privacy is not for me, it's for those rich
  women'': Performative Privacy Practices on Mobile Phones by Women in South
  Asia}. In \bibinfo{booktitle}{\emph{Fourteenth Symposium on Usable Privacy
  and Security (SOUPS 2018)}}. \bibinfo{pages}{127--142}.
\newblock


\bibitem[\protect\citeauthoryear{Scott, Graham, Wallace, Hancock, and
  Nacenta}{Scott et~al\mbox{.}}{2015}]%
        {scott2015local}
\bibfield{author}{\bibinfo{person}{Stacey~D Scott},
  \bibinfo{person}{TC~Nicholas Graham}, \bibinfo{person}{James~R Wallace},
  \bibinfo{person}{Mark Hancock}, {and} \bibinfo{person}{Miguel Nacenta}.}
  \bibinfo{year}{2015}\natexlab{}.
\newblock \showarticletitle{``Local Remote'' Collaboration: Applying Remote
  Group AwarenessTechniques to Co-located Settings}. In
  \bibinfo{booktitle}{\emph{Proceedings of the 18th ACM Conference Companion on
  Computer Supported Cooperative Work \& Social Computing}}.
  \bibinfo{pages}{319--324}.
\newblock


\bibitem[\protect\citeauthoryear{{Snap Inc.}}{{Snap Inc.}}{2017}]%
        {lensstudio}
\bibfield{author}{\bibinfo{person}{{Snap Inc.}}}
  \bibinfo{year}{2017}\natexlab{}.
\newblock \bibinfo{title}{Lens Studio}.
\newblock
\newblock
\urldef\tempurl%
\url{https://lensstudio.snapchat.com/}
\showURL{%
\tempurl}
\newblock
\shownote{(Accessed on 02/04/2023).}


\bibitem[\protect\citeauthoryear{{Snap Inc.}}{{Snap Inc.}}{2021}]%
        {connectedlenses}
\bibfield{author}{\bibinfo{person}{{Snap Inc.}}}
  \bibinfo{year}{2021}\natexlab{}.
\newblock \bibinfo{title}{Connected Lenses Overview}.
\newblock
\newblock
\urldef\tempurl%
\url{https://lensstudio.snapchat.com/guides/connected-lenses/connected-lenses-overview/}
\showURL{%
\tempurl}
\newblock
\shownote{(Accessed on 02/04/2023).}


\bibitem[\protect\citeauthoryear{Stepanova, Desnoyers-Stewart, Pasquier, and
  Riecke}{Stepanova et~al\mbox{.}}{2020}]%
        {stepanova2020jel}
\bibfield{author}{\bibinfo{person}{Ekaterina~R Stepanova},
  \bibinfo{person}{John Desnoyers-Stewart}, \bibinfo{person}{Philippe
  Pasquier}, {and} \bibinfo{person}{Bernhard~E Riecke}.}
  \bibinfo{year}{2020}\natexlab{}.
\newblock \showarticletitle{JeL: Breathing together to connect with others and
  nature}. In \bibinfo{booktitle}{\emph{Proceedings of the 2020 ACM Designing
  Interactive Systems Conference}}. \bibinfo{pages}{641--654}.
\newblock


\bibitem[\protect\citeauthoryear{Sullivan, McCoy, Hendricks, and
  Williams}{Sullivan et~al\mbox{.}}{2018}]%
        {Loominar90:online}
\bibfield{author}{\bibinfo{person}{Anne Sullivan},
  \bibinfo{person}{Joshua~Allen McCoy}, \bibinfo{person}{Sarah Hendricks},
  {and} \bibinfo{person}{Brittany Williams}.} \bibinfo{year}{2018}\natexlab{}.
\newblock \showarticletitle{Loominary: Crafting Tangible Artifacts from Player
  Narrative}. In \bibinfo{booktitle}{\emph{Proceedings of the Twelfth
  International Conference on Tangible, Embedded, and Embodied Interaction (TEI
  '18)}}. \bibinfo{pages}{443--450}.
\newblock


\bibitem[\protect\citeauthoryear{Sun, De~Oliveira, and Lewandowski}{Sun
  et~al\mbox{.}}{2017}]%
        {sun2017challenges}
\bibfield{author}{\bibinfo{person}{Emily Sun}, \bibinfo{person}{Rodrigo
  De~Oliveira}, {and} \bibinfo{person}{Joshua Lewandowski}.}
  \bibinfo{year}{2017}\natexlab{}.
\newblock \showarticletitle{Challenges on the journey to co-watching YouTube}.
  In \bibinfo{booktitle}{\emph{Proceedings of the 2017 ACM Conference on
  Computer Supported Cooperative Work and Social Computing}}.
  \bibinfo{pages}{783--793}.
\newblock


\bibitem[\protect\citeauthoryear{Surale, Tham, Smith, and Vaish}{Surale
  et~al\mbox{.}}{2022}]%
        {ARcall:online}
\bibfield{author}{\bibinfo{person}{Hemant~Bhaskar Surale},
  \bibinfo{person}{Yu~Jiang Tham}, \bibinfo{person}{Brian~A. Smith}, {and}
  \bibinfo{person}{Rajan Vaish}.} \bibinfo{year}{2022}\natexlab{}.
\newblock \showarticletitle{ARcall: Real-Time AR Communication using
  Smartphones and Smartglasses}. In \bibinfo{booktitle}{\emph{Augmented Humans
  2022 (AHs 2022)}}. \bibinfo{pages}{1--19}.
\newblock


\bibitem[\protect\citeauthoryear{Torres}{Torres}{2019}]%
        {KindWord16:online}
\bibfield{author}{\bibinfo{person}{Alan Torres}.}
  \bibinfo{year}{2019}\natexlab{}.
\newblock \bibinfo{title}{Kind Words: A Game Awards Nominee About The Kindness
  of Strangers}.
\newblock
  \bibinfo{howpublished}{\url{https://www.digitaltrends.com/gaming/kind-words-game-awards-2019/}}.
\newblock
\newblock
\shownote{(Accessed on 02/04/2023).}


\bibitem[\protect\citeauthoryear{Trindade, Sousa, Hart, Vieira, Rodrigues, and
  Fran{\c{c}}a}{Trindade et~al\mbox{.}}{2015}]%
        {trindade2015purrfect}
\bibfield{author}{\bibinfo{person}{Rui Trindade}, \bibinfo{person}{Micaela
  Sousa}, \bibinfo{person}{Cristina Hart}, \bibinfo{person}{N{\'a}dia Vieira},
  \bibinfo{person}{Roberto Rodrigues}, {and} \bibinfo{person}{Jo{\~a}o
  Fran{\c{c}}a}.} \bibinfo{year}{2015}\natexlab{}.
\newblock \showarticletitle{Purrfect crime: Exploring animal computer
  interaction through a digital game for humans and cats}. In
  \bibinfo{booktitle}{\emph{Proceedings of the 33rd Annual ACM Conference
  Extended Abstracts on Human Factors in Computing Systems}}.
  \bibinfo{pages}{93--96}.
\newblock


\bibitem[\protect\citeauthoryear{Tsai, Hsu, Chang, Huang, Ho, and LaRose}{Tsai
  et~al\mbox{.}}{2019}]%
        {tsai2019high}
\bibfield{author}{\bibinfo{person}{Hsin-yi~Sandy Tsai},
  \bibinfo{person}{Pei-Jung Hsu}, \bibinfo{person}{Chih-Ling Chang},
  \bibinfo{person}{Chun-Cheng Huang}, \bibinfo{person}{Hsin-Fang Ho}, {and}
  \bibinfo{person}{Robert LaRose}.} \bibinfo{year}{2019}\natexlab{}.
\newblock \showarticletitle{High tension lines: Negative social exchange and
  psychological well-being in the context of instant messaging}.
\newblock \bibinfo{journal}{\emph{Computers in Human Behavior}}
  \bibinfo{volume}{93} (\bibinfo{year}{2019}), \bibinfo{pages}{326--332}.
\newblock


\bibitem[\protect\citeauthoryear{Turkle}{Turkle}{2011}]%
        {alonetogether2011}
\bibfield{author}{\bibinfo{person}{Sherry Turkle}.}
  \bibinfo{year}{2011}\natexlab{}.
\newblock \bibinfo{booktitle}{\emph{Alone Together: Why We Expect More from
  Technology and Less from Each Other.}}
\newblock \bibinfo{publisher}{New York: Basic Books}.
\newblock


\bibitem[\protect\citeauthoryear{Turkle}{Turkle}{2017}]%
        {turkle2017alone}
\bibfield{author}{\bibinfo{person}{Sherry Turkle}.}
  \bibinfo{year}{2017}\natexlab{}.
\newblock \bibinfo{booktitle}{\emph{Alone Together: Why We Expect More from
  Technology and Less from Each Other} (\bibinfo{edition}{third} ed.)}.
\newblock \bibinfo{publisher}{Hachette UK}.
\newblock


\bibitem[\protect\citeauthoryear{Umbelino, Ta, Blake, Truong, Luo, and
  Dow}{Umbelino et~al\mbox{.}}{2019}]%
        {umbelino2019prototeams}
\bibfield{author}{\bibinfo{person}{Gustavo Umbelino}, \bibinfo{person}{Vivian
  Ta}, \bibinfo{person}{Samuel Blake}, \bibinfo{person}{Eric Truong},
  \bibinfo{person}{Amy Luo}, {and} \bibinfo{person}{Steven Dow}.}
  \bibinfo{year}{2019}\natexlab{}.
\newblock \showarticletitle{ProtoTeams: Supporting Small Group Interactions in
  Co-Located Crowds}. In \bibinfo{booktitle}{\emph{Conference Companion
  Publication of the 2019 on Computer Supported Cooperative Work and Social
  Computing}}. \bibinfo{pages}{392--397}.
\newblock


\bibitem[\protect\citeauthoryear{Van~House}{Van~House}{2009}]%
        {van2009collocated}
\bibfield{author}{\bibinfo{person}{Nancy~A Van~House}.}
  \bibinfo{year}{2009}\natexlab{}.
\newblock \showarticletitle{Collocated photo sharing, story-telling, and the
  performance of self}.
\newblock \bibinfo{journal}{\emph{International Journal of Human-Computer
  Studies}} \bibinfo{volume}{67}, \bibinfo{number}{12} (\bibinfo{year}{2009}),
  \bibinfo{pages}{1073--1086}.
\newblock


\bibitem[\protect\citeauthoryear{Vetere, Gibbs, Kjeldskov, Howard, Mueller,
  Pedell, Mecoles, and Bunyan}{Vetere et~al\mbox{.}}{2005}]%
        {vetere2005}
\bibfield{author}{\bibinfo{person}{Frank Vetere}, \bibinfo{person}{Martin~R.
  Gibbs}, \bibinfo{person}{Jesper Kjeldskov}, \bibinfo{person}{Steve Howard},
  \bibinfo{person}{Florian~'Floyd' Mueller}, \bibinfo{person}{Sonja Pedell},
  \bibinfo{person}{Karen Mecoles}, {and} \bibinfo{person}{Marcus Bunyan}.}
  \bibinfo{year}{2005}\natexlab{}.
\newblock \showarticletitle{Mediating Intimacy: Designing Technologies to
  Support Strong-Tie Relationships}. In \bibinfo{booktitle}{\emph{Proceedings
  of the SIGCHI Conference on Human Factors in Computing Systems}} (Portland,
  Oregon, USA) \emph{(\bibinfo{series}{CHI '05})}.
  \bibinfo{publisher}{Association for Computing Machinery},
  \bibinfo{address}{New York, NY, USA}, \bibinfo{pages}{471–480}.
\newblock
\showISBNx{1581139985}
\urldef\tempurl%
\url{https://doi.org/10.1145/1054972.1055038}
\showDOI{\tempurl}


\bibitem[\protect\citeauthoryear{Weil}{Weil}{2017}]%
        {weil2017three}
\bibfield{author}{\bibinfo{person}{Andrew Weil}.}
  \bibinfo{year}{2017}\natexlab{}.
\newblock \showarticletitle{Three Breathing Exercises and Techniques}.
\newblock
  \bibinfo{howpublished}{\url{https://www.drweil.com/health-wellness/body-mind-spirit/stress-anxiety/breathing-three-exercises/}}.
\newblock \bibinfo{journal}{\emph{DrWeil.com}} (\bibinfo{year}{2017}).
\newblock


\bibitem[\protect\citeauthoryear{Westerlaken and Gualeni}{Westerlaken and
  Gualeni}{2014}]%
        {westerlaken2014felino}
\bibfield{author}{\bibinfo{person}{Michelle Westerlaken} {and}
  \bibinfo{person}{Stefano Gualeni}.} \bibinfo{year}{2014}\natexlab{}.
\newblock \showarticletitle{Felino: the philosophical practice of making an
  interspecies video game}.
\newblock \bibinfo{journal}{\emph{The Game Philosophy Network}}
  (\bibinfo{year}{2014}).
\newblock


\bibitem[\protect\citeauthoryear{Wingrave, Rose, Langston, and
  LaViola~Jr}{Wingrave et~al\mbox{.}}{2010}]%
        {wingrave2010early}
\bibfield{author}{\bibinfo{person}{Chadwick~A Wingrave},
  \bibinfo{person}{Jeremy Rose}, \bibinfo{person}{Todd Langston}, {and}
  \bibinfo{person}{Joseph~J LaViola~Jr}.} \bibinfo{year}{2010}\natexlab{}.
\newblock \showarticletitle{Early explorations of CAT: canine amusement and
  training}.
\newblock In \bibinfo{booktitle}{\emph{CHI'10 Extended Abstracts on Human
  Factors in Computing Systems}}. \bibinfo{pages}{2661--2670}.
\newblock


\bibitem[\protect\citeauthoryear{Wu, Orlosky, Kiyokawa, and Takemura}{Wu
  et~al\mbox{.}}{2016}]%
        {wuemotopet}
\bibfield{author}{\bibinfo{person}{Qifan Wu}, \bibinfo{person}{Jason Orlosky},
  \bibinfo{person}{Kiyoshi Kiyokawa}, {and} \bibinfo{person}{Haruo Takemura}.}
  \bibinfo{year}{2016}\natexlab{}.
\newblock \showarticletitle{EmotoPet: Exploring Emotion with an
  Environment-sensing Virtual Pet}.
\newblock  (\bibinfo{year}{2016}).
\newblock


\bibitem[\protect\citeauthoryear{Zhao, Bennett, Benko, Cutrell, Holz, Morris,
  and Sinclair}{Zhao et~al\mbox{.}}{2018}]%
        {canetrol34:online}
\bibfield{author}{\bibinfo{person}{Yuhang Zhao}, \bibinfo{person}{Cynthia~L.
  Bennett}, \bibinfo{person}{Hrvoje Benko}, \bibinfo{person}{Edward Cutrell},
  \bibinfo{person}{Christian Holz}, \bibinfo{person}{Meredith~Ringel Morris},
  {and} \bibinfo{person}{Mike Sinclair}.} \bibinfo{year}{2018}\natexlab{}.
\newblock \showarticletitle{Enabling People with Visual Impairments to Navigate
  Virtual Reality with a Haptic and Auditory Cane Simulation}. In
  \bibinfo{booktitle}{\emph{Proceedings of the 2018 CHI Conference on Human
  Factors in Computing Systems}}. \bibinfo{pages}{1--14}.
\newblock


\end{thebibliography}



\end{document}